\documentclass[10pt.draft]{article}
\usepackage{amssymb}
\usepackage{graphicx}
\usepackage{amsmath}
\usepackage{acronym}

\input{tcilatex}

\begin{document}

\bigskip

\ {\huge Traces of Mirror Symmetry in Nature}

$\ \ \ \ \ \ \ \ \ \ \ \ \ \ \ \ \ \ \ \ \ \ \ \ \ \ \ \ \ \ \ \ \ $

\ \ \ \ \ \ \ \ \ \ \ \ \ \ \ \ \ \ \ \ \ \ \ \ \ \ \ A.L. Kholodenko%
\footnote{%
E-mail address: string@clemson.edu}

\textit{375 H.L.Hunter Laboratories, Clemson University, Clemson, }

\textit{SC} 29634-0973, USA

\bigskip

In this work we discuss the place of Veneziano amplitudes (the precursor of
string models) and their generalizations in the Regge theory of high energy
physics scattering processes. We emphasize that mathematically such
amplitudes and their extensions can be interpreted in terms of the Laplace
(respectively, multiple Laplace) transform(s) of the generating function for
the Ehrhart polynomial associated with some integral polytope $\mathcal{P}$
(specific for each scattering process). Following works by Batyrev and Hibi
\ to each such polytope $\mathcal{P}$ it is possible to associate another
(mirror) polytope $\mathcal{P}^{\prime }$. \ For this to happen, it is
necessary to impose some conditions on $\mathcal{P}$ and, hence, on the
generating function for $\mathcal{P}$. Since each of these polytopes is in
fact encodes some projective toric variety, this information is used for
development of new symplectic and supersymmetric models reproducing the
Veneziano and generalized Veneziano amplitudes. General ideas are
illustrated on classical example of the pion-pion scattering for which the
existing experimental data can be naturally explained with help of mirror
symmetry arguments.

\ 


\pagebreak

\ \ \ 

\section{\protect\bigskip Introduction}

\ 

\subsection{Brief history of the Veneziano amplitudes}

\bigskip

As is well known, the origins of modern string theory can be traced back to
the 4-particle scattering amplitude $A(s,t,u)$ \ postulated by Veneziano in
1968 [1]. Up to a common constant factor, it is given by 
\begin{equation}
A(s,t,u)=V(s,t)+V(s,u)+V(t,u),  \tag{1}
\end{equation}%
where 
\begin{equation}
V(s,t)=\int\limits_{0}^{1}x^{-\alpha (s)-1}(1-x)^{-\alpha (t)-1}dx\equiv
B(-\alpha (s),-\alpha (t))  \tag{2}
\end{equation}%
is the Euler beta function and $\alpha (x)$ is the Regge trajectory usually
written as $\alpha (x)=\alpha (0)+\alpha ^{\prime }x$ with $\alpha (0)$ and $%
\alpha ^{\prime }$ being the Regge slope and intercept, respectively. In the
case of space-time metric with signature $\{-,+,+,+\}$ the Mandelstam
variables $s$, $t$ and $u$ entering the Regge trajectory are defined by [2] 
\begin{equation}
s=-(p_{1}+p_{2})^{2};\text{ }t=-(p_{2}+p_{3})^{2};\text{ }%
u=-(p_{3}+p_{1})^{2}.  \tag{3}
\end{equation}%
The 4-momenta $p_{i}$\ are constrained by the energy-momentum conservation
law leading to relation between the Mandelstam variables: 
\begin{equation}
s+t+u=\sum\limits_{i=1}^{4}m_{i}^{2}.  \tag{4}
\end{equation}%
Veneziano [1] noticed\footnote{%
To get our Eq.(5) from Eq.(7) of Veneziano paper, it is sufficient to notice
that his $1-\alpha (s)$ corresponds to ours -$\alpha (s).$} that to fit
experimental data the Regge trajectories should obey the constraint 
\begin{equation}
\alpha (s)+\alpha (t)+\alpha (u)=-1  \tag{5}
\end{equation}%
consistent with Eq.(4) in view of the definition of $\alpha (s).$ The
Veneziano condition, Eq.(5),\ can be rewritten in a more general form.
Indeed, let $m,n,l$ be some integers such that $\alpha (s)m+\alpha
(t)n+\alpha (u)l=0$. Then by adding this equation to Eq.(5) we obtain, $%
\alpha (s)\tilde{m}+\alpha (t)\tilde{n}+\alpha (u)\tilde{l}=-1,$ or more
generally, $\alpha (s)\tilde{m}+\alpha (t)\tilde{n}+\alpha (u)\tilde{l}+%
\tilde{k}\cdot 1=0.$ Both equations have been studied extensively in the
book by Stanley [3] from the point of view of commutative algebra,
polytopes, toric varieties, invariants of finite groups, etc. Although this
observation is \ entirely sufficient for restoration of the underlying
physical model(s) reproducing these amplitudes, development of
string-theoretic models reproducing such amplitudes proceeded historically
quite differently. In this work, we abandon these more traditional
approaches in favour of taking the full advantage of combinatorial ideas
presented in Ref.[3]. This allows us to obtain models reproducing Veneziano
amplitudes \ which are markedly different from those known in traditional
string-theoretic literature.

In 1967-a year before Veneziano's paper was published- the paper [4] by
Chowla and Selberg appeared relating Euler's beta function to the periods of
elliptic integrals. The result by Chowla and Selberg was generalized by
Andre Weil whose two influential papers [5,6] brought into the picture the
periods of Jacobians of the Abelian varieties, Hodge rings, etc. Being
motivated by these papers, Benedict Gross wrote a paper [7] in which the
beta function appears as period associated with the differential form
\textquotedblright living\textquotedblright\ on the Jacobian of the Fermat
curve. His results as well as those by Rohrlich (placed in the appendix to
Gross paper) have been subsequently documented in the book by Lang [8].
Perhaps, because in the paper by Gross the multidimensional extension of
beta function was considered only briefly, e.g.[7],p.207, the computational
details were not provided. These details can be found in our recently
published papers, Refs.[9,10,11]. To obtain the multidimensional extension
of beta function as period integral, following the logic of papers by Gross
and Deligne [12], one needs to replace the Fermat curve by the Fermat
hypersufrace, to embed it into the complex projective space, and to treat it
as K\"{a}hler manifold. The differential forms living on such manifold are
associated with the periods of Fermat hypersurface. Physical considerations
require this K\"{a}hler manifold to be of the Hodge type. In his lecture
notes [12] Deligne noticed that the Hodge theory needs some essential
changes (e.g. mixed Hodge structures, etc.) if the Hodge-K\"{a}hler
manifolds possess singularities. Such modifications may be needed upon
development of our formalism. A monograph by Carlson et al, Ref.[13],
contains an up to date exaustive information regarding such modifications,
etc. Fortunately, to obtain the multiparticle Veneziano amplitudes these
complications are not essential. In Ref.[10] we demonstrated that the period
integrals living on Fermat hypersurfaces, when properly interpreted, provide
the tachyon-free (Veneziano-like) multiparticle amplitudes whose particle
spectrum reproduces those known for both the open and closed bosonic
strings. Naturally, the question arises: If this is so, then what kind of
models are capable of reproducing such amplitudes ?\ In this paper we would
like to discuss some combinatorial properties of the Veneziano (and
Veneziano-like) amplitudes sufficient for reproducing \ at least two of such
models: symplectic and supersymmetric. Mathematically, the results presented
below are in accord with those by Vergne [14] whose work does not contain
practical applications. Before studying these models, we would like to make
some comments about the place of Veneziano amplitudes and, hence, of
whatever models associated with these amplitudes, within the Regge formalism
\ developed for description of scattering processes in high energy physics.
This is accomplished in the next subsection.

\ 

\subsection{The Regge theory, theorem by Froissart, quantum gravity and the
standard model}

\bigskip

As is well known, all information in particle physics is obtainable through
proper interpretation of the scattering data.The optical theorem (see below)
allows one to connect the imajinary part of the scattering amplitude with
the total crossection $\sigma .$ By measuring this crossection
experimentally one can obtain some information about the scattering
amplitudes. Additional useful information can be obtained by collecting data
for differential crossections, by using the dispersion relations, etc.[15].\
\ There is an unproven common belief that in the limit of high energies 
\textsl{all} scattering processes are adequately described by the Regge
theory [16, 17]. The Veneziano amplitude \textsl{by design} is Regge
behaving [1]. To our knowledge, the proof that in the limit of high energies
scattering amplitudes are Regge behaving had been obtained \ only for some
special cases [16,17], including that of QCD [18]. Since, irrespective to
their mathematical nature, all string theories are based on this ( generally
unproven!) belief of the validity of the Regge theory, they can be as much
trusted (even if totally correct mathematically !) as can be the Regge
theory.

In the Regge theory the experimental data are presented using the
Chew-Frautchi (C-F) plot, Ref.[16], pp. 144-145. On this plot one plots the
Regge trajectories. Such trajectories relate particles with the same
internal quantum numbers but with different spin (or angular momentum). From
the standard string textbook, Ref.[2], it is known that for the open bosonic
string the Regge trajectory is given by $\alpha (s)=\alpha (0)+\alpha
^{\prime }s$ \ (in accord with Eq.(2) above). It is important though that $%
\alpha (0)=1$ and $\alpha ^{\prime }=1/2$ for the \textsl{open} string while 
$\alpha (0)=2$ and $\alpha ^{\prime }=1/4$ for the \textsl{closed} string.
In known string-theoretic formulations the numerical values of these
parameters \textsl{cannot} be adjusted to fit the available experimental
data since their values are deeply connected with the existing
string-theoretic formalism [2] and, hence, are not readily adjustable. In
the meantime, for high energies currently available it is known, e.g. read
Ref.[15], p. 41, that $\alpha (s)=0.7+0.8s$ or \ $\alpha (s)=0.44+0.92s$ \
for typical Regge tragectories. Claims made by some string theoreticians
that the available range of high energies is not sufficient to test the
predictions provided by the existing string theories cannot be justified
because of the following.

One of the major reasons for development of string theory, according to
Ref.[2], lies in developing of consitent theory of quantum gravity. Indeed,
in the case of closed bosonic string the massless (i.e. $s=0$) spin two
graviton occurs in the string spectrum only if $\alpha (0)=2.$This fact
alone fixes the value of the Regge intercept $\alpha (0)$ on the C-F plot to
its value : $\alpha (0)=2.$ As plausible as it is, such an identification
creates some major problems.

Indeed, in the case of $2\rightarrow 2$ scattering process the total cross
section for the elastic scattering in $s$-channel (in view of the optical
theorem, e.g. see Ref.[15], p. 47) is given by 
\begin{equation}
\sigma (s)\sim s^{-1}\func{Im}A(s,t=0),  \tag{6}
\end{equation}%
where the scattering amplitude $A(s,t)$ is either postulated (as in the case
of Veneziano amplitude) or determined from some model (e.g.\ the standard
string model [2], etc.). The above expression is valid rigorously at \textsl{%
any} energy. In the limit $s\rightarrow \infty $ the Regge theory provides
the estimate for this exact result : 
\begin{equation}
\sigma (s)=cs^{\alpha (t=0)-1},  \tag{7}
\end{equation}%
where $c$ is some constant. As it is with all processes described by the
Regge theory [15-17], physically this result means the following : the
analytical behaviour of the amplitude for elastic scattering in the $s$%
-channel is controlled (through the exponent in Eq.(7)) by the resonance in $%
t-$channel. In particular, if the resonance is caused by the graviton this
leads the total crossection to behave as: $\sigma (s)=cs.$ Unfortunately,
the obtained result violates the theorem by Froissart. It can be stated as
follows (e.g. see Ref.[16], p.53) :

\ 

\bigskip

\textbf{Theorem 1.1. (}Froissart\textbf{) }In the high energy limit : $%
s\rightarrow \infty $ the total crossection $\sigma (s)$ in s-channel is
bounded by $\sigma (s)_{s\rightarrow \infty }\leq const\log (s/s_{0})$ \
where $s_{0}$ is some (prescribed) energy scale.

\bigskip

Evidently, even if the current efforts (based on commonly accepted
formalism) to construct mathematically meaningful string/brane theory
eventually might succeed, such a theory will contradict the Froissart
theorem for reasons just described. Hence, either this theorem is incorrect
and should be reconsidered or the underlying assumptions of string theory
regarding gravitons are incorrect.

\ 

\textbf{Remark 1.2}. The way out from this situation was recently developed
in our recent work, Ref.[18], where new equivalence principle for gravity is
proposed based on known rigorous mathematical results. This new equivalence
principle has major implications for the standard model of particle physics
[19]. Since \ physical predictions based this model are in agreement with
the Froissart theorem already, the results of Ref.[18] effectively convert
the existing standard model into a unified field theory accounting for all
four types of known fundamental interactions and being manifestly
renormalizable and gauge-invariant.

\ 

Incidentally, the intercept $\alpha (0)=1$ for the open string theory does
have some physical significance. Indeed, in this case use of Eq.(7) produces 
$\sigma (s)=c^{\prime }$ where $c^{\prime }$ is yet another constant. Such
high energy begaviour is typical for the \textit{pomeron}-a hypothetical
particle like object predicted by Pomeranchuk- still undinscovered [15,17].
Additional ramifications of Pomeranchuk's work have lead to the prediction
of the companion of the pomeron-the \textit{oddero}n [20].

In addition to the difficulty with the Froissart theorem, just described,
the existing string-theoretic models suffer from several no less serious
drawbacks. For instance, the Regge theory in general and the Veneziano
amplitude (a precursor of the string model) in particular states that in
addition to the leading (\textit{parent}) \ Regge trajectory there should be 
\textit{countable infinity} of \textit{daughter} trajectories-all lying
below the parent trajectory. Nowhere in string-theoretic literature were we
able to find a mention or an explanation of this fact. Experimentally,
however, typically for each parent trajectory there are only few daughter
trajectories. In this work we shall provide a plausible theoretical
explanation of this fact based on the mirror symmetry arguments. We \ would
like to emphasize that since the models reproducing Veneziano amplitudes
discussed below differ from those commonly discussed in string-theoretic
literature, the numerical values for the slope $\alpha ^{\prime }$ and the
intercept $\alpha (0)$ of the Regge trajectories can be readily adjusted to
fit the experimental data. This is in accord with the original work by
Veneziano [1] where no restrictions on the slope and intercept were imposed.

\subsection{Organization of the rest of this paper}

The rest of this work is organized as follows. Section 2 begins with some
facts revealing the combinatorial nature of Veneziano amplitudes. This is
achieved by connecting them with generating function for the Ehrhart
polynomial whose properties are described in some detail in the same
section. Such a polynomial counts the number of points inside the rational
polytope (i.e. polytope whose vertices are located at the nodes of the
regular $k-$dimensional lattice) and at its boundaries (faces). In the
present case the polytope is a regular simplex which is a deformation
retract for the Fermat-type (hyper) surface living in the complex projective
space [9,10]. Next, using general properties of generating functions for the
Ehrhart polynomials for the rational polytopes we discuss possible
generalizations of the Veneziano amplitudes for polytopes other than a
simplex. This allows us to use some results by Batyrev [22, 23] and Hibi 
\textbf{[}24] in order to introduce the mirror symmetry considerations
enabling us to exclude the countable infinity of daugher trajectories on the
C-F plot using mirror symmetry arguments.\ General ideas are illustraded on
the classical example of the pion-pion scattering [25] for which the
existing experimental data can be naturally explained with help of mirror
symmetry arguments. Next, in Section 3 we begin our reconstruction of the
models reproducing Veneziano and the generalized Veneziano amplitudes. It is
facilitated by known connections between the polytopes and dynamical systems
[14,26]. Development of these connections is proceeds through Sections 2-4
where we find the corresponding quantum mechanical system whose ground state
is degenerate with degeneracy factor being identified with the Ehrhart
polynomial. The obtained final result is in accord with that earlier
obtained by Vergne [14] whose work does not contain any physical
applications. In Section 5 the generating function for the Ehrhart
polynomial is reinterpreted in terms of the Poincar$e^{\prime }$ polynomial.
Such a polynomial is used, for instance, in the theory of invariants of
finite (pseudo)reflection groups [3,27]. Obtained indentification reveals
the topological and group-theoretic nature of the Veneziano amplitudes. To
strengthen this point of view, we use some results by Atiyah and Bott [28]
inspired by earlier work by Witten [29] on supersymmetric quantum mechanics.
They allow us to think about the Veneziano amplitudes using the therminology
of intersection theory [30]. This is consistent with earlier mentioned
interpretation of these amplitudes in terms of periods of the Fermat
(hyper)surface [9,10]. It also makes computation of these amplitudes
analogous to those for the Witten -Kontsevich model [31, 32], whose
refinements can be found in our earlier work, Ref.[33]. For the sake of
space, in this work we do not develop these connections with the Witten-
Kontsevich model any further. Interested reader may find them in Ref.[34].
Instead, we discuss the supersymmetric model associated with symplectic
model described earlier and treat it with help of the Lefshetz isomorphism
theorem. This allows us to look at the problem of computation of the
spectrum for such a model from the point of view of the theory of
representations of the complex semisimple Lie algebras. Using some results
by Serre [35] and Ginzburg [36] we demonstrate that the ground state for
such finite dimensional supersymmetric quantum mechanical model is
degenerate with degeneracy factor coinciding with the Erhardt polynomial.
This result is consistent with that obtained in Section 4 by different
methods.

\section{The extended Veneziano amplitudes, the Ehrhart polynomial and
mirror symmetry}

\subsection{\protect\bigskip Combinatorics of the Veneziano amplitudes}

In view of Eq.(2), consider an identity taken from [37],

\begin{align}
\frac{1}{(1-tz_{0})\cdot \cdot \cdot (1-tz_{k})}& =(1+tz_{0}+\left(
tz_{0}\right) ^{2}+...)\cdot \cdot \cdot (1+tz_{n}+\left( tz_{n}\right)
^{2}+...)  \notag \\
& =\sum\limits_{n=0}^{\infty
}(\sum\limits_{k_{0}+...+k_{k}=n}z_{0}^{k_{0}}\cdot \cdot \cdot
z_{k}^{k_{k}})t^{n}.  \tag{8}
\end{align}%
When $z_{0}=...=z_{k}=1,$the inner sum in the last expression provides the
total number of monomials of the type $z_{0}^{k_{0}}\cdot \cdot \cdot
z_{k}^{k_{k}}$ with $k_{0}+...+k_{k}=n$. The total number of such monomials
is given by the binomial coefficient\footnote{%
The reason for displaying 3 different forms of the same combinatorial factor
will be explained shortly below.} 
\begin{equation}
p(k,n)=\frac{\left( k+n\right) !}{k!n!}=\frac{(n+1)(n+2)\cdot \cdot \cdot
(n+k)}{k!}=\frac{(k+1)(k+2)\cdot \cdot \cdot (k+n)}{n!}.  \tag{9}
\end{equation}%
For this special case Eq.(8) is converted to a useful expansion, 
\begin{equation}
P(k,t)\equiv \frac{1}{\left( 1-t\right) ^{k+1}}=\sum\limits_{n=0}^{\infty
}p(k,n)t^{n}.  \tag{10}
\end{equation}%
In view of the integral representation of the beta function given by Eq.(2),
\ we replace $k+1$ by $\alpha (s)+1$ in Eq.(10) and use it in the beta
function representation of the amplitude $V(s,t)$. Straightforward
calculation produces the following known in string theory result [2]: 
\begin{equation}
V(s,t)=-\sum\limits_{n=0}^{\infty }p(\alpha (s),n)\frac{1}{\alpha (t)-n}. 
\tag{11}
\end{equation}%
The r.h.s. of Eq.(11) is effectively the Laplace transform of the generating
function, Eq.(10).\textsl{\ Such generating function can be interpreted as a
partition function in the sence of} \textsl{statistical mechanics}.

\textsl{The purpose of this work is to demonstrate that such an
interpretation is not merely a conjecture} and, in view of this, \textsl{to
find the statisical mechanical/quantum model whose partition function is
given by} Eq.(10).

Our arguments are not restricted to the 4-particle amplitude. Indeed, as we
argued earlier [10,11], the multidimensional extension of Euler's beta
function producing murtiparticle Veneziano amplitudes (upon symmetrization
analogous to the 4-particle case) is given by the following integral
attributed to Dirichlet 
\begin{equation}
\mathcal{D}(x_{1},...,x_{k})=\int \int\limits_{\substack{ u_{1}\geq
0,...,u_{k}\geq 0  \\ u_{1}\text{ }+\cdot \cdot \cdot +u_{k}\leq 1}}\unit{u}%
_{1}^{x_{1}-1}\unit{u}_{2}^{x_{2}-1}...\unit{u}%
_{k}^{x_{k}-1}(1-u_{1}-...-u_{k})^{x_{k+1}-1}du_{1}...du_{k}.  \tag{12}
\end{equation}%
In this integral let $t=u_{1}+...+u_{k}$. This allows us to use already
familiar expansion Eq.(10). In addition, the following identity 
\begin{equation}
t^{n}=(u_{1}+...+u_{k})^{n}=\sum\limits_{n=(n_{1},...,n_{k})}\frac{n!}{%
n_{1}!n_{2}!...n_{k}!}u_{1}^{n_{1}}\cdot \cdot \cdot u_{k}^{n_{k}}  \tag{13}
\end{equation}%
with restriction $n=n_{1}+...+n_{k}$ is of importance as well. This type of
identity was used earlier in our work on Kontsevich-Witten model [33].
Moreover, from the same paper it follows that the above result can be
presented as well in the alternative useful form: 
\begin{equation}
(u_{1}+...+u_{k})^{n}=\sum\limits_{\lambda \vdash k}f^{\lambda }S_{\lambda
}(u_{1},...,u_{k}),  \tag{14}
\end{equation}%
where the Schur polynomial $S_{\lambda }$ is defined by 
\begin{equation}
S_{\lambda }(u_{1},...,u_{k})=\sum\limits_{n=(n_{1},...,n_{k})}K_{\lambda
,n}u_{1}^{n_{1}}\cdot \cdot \cdot u_{k}^{n_{k}}  \tag{15}
\end{equation}%
with coefficients $K_{\lambda ,n}$ known as Kostka numbers, $\ f^{\lambda }$
being the number of standard Young tableaux of shape $\lambda $ and the
notation $\lambda \vdash k$ meaning that $\lambda $ is partition of $k$.
Through such a connection with Schur polynomials one can develop connections
with the Kadomtsev-Petviashvili (KP) hierarchy of nonlinear exactly
integrable systems on one hand\ and with the theory of Schubert varieties on
another. \ Although details \ can be found in our earlier publications
[33,11], in this work we shall discuss these issues a bit further in Section%
\textbf{\ }5\textbf{. }Use of Eq.(13) in (12) produces, after performing the
multiple Laplace transform, the following part of the multiparticle
Veneziano amplitude 
\begin{equation}
A(1,...k)=\frac{\Gamma _{n_{1}...n_{k}}(\alpha (s_{k+1}))}{(\alpha
(s_{1})-n_{1})\cdot \cdot \cdot (\alpha (s_{k})-n_{k})}.  \tag{16}
\end{equation}%
Even though the residue $\Gamma _{n_{1}...n_{k}}(\alpha (s_{k+1}))$ contains
all the combinatorial factors, the obtained result should still be
symmetrized (in accord with the 4-particle case considered by Veneziano) in
order to obtain the full murtiparticle Veneziano amplitude. Since in the
above general multiparticle case the same expansion, Eq.(10), was used,\ for
the sake of space it is sufficient to focus on the 4-particle amplitude
only. This task is reduced to further study of the expansion given by
Eq.(10). Such an expansion can be looked upon from several different angles.
For instance, we have mentioned already that it can be interpreted as a
partition function. In addition, it is the generating function for the
Ehrhart polynomial. The combinatorial factor $p(k,n)$ defined in Eq.(9) is
the simplest example of the Ehrhart polynomial. Evidently, it can be written
formally as 
\begin{equation}
p(k,n)=a_{n}k^{n}+a_{n-1}k^{n-1}+\cdot \cdot \cdot +a_{0}.  \tag{17}
\end{equation}

\subsection{\protect\bigskip Some facts about the Ehrhart polynomials}

A type of expansion given by Eq.(17) is typical \ for \textsl{all} \
Ehrhart-type polynomials. Indeed, let $\mathcal{P}$ be \textit{any} convex
rational polytope that is the polytope whose vertices are located at the
nodes of some $n-$dimensional \textbf{Z}$^{n}$ lattice. Then, the Ehrhart
polynomial \ for the inflated polytope $\mathcal{P}$ (with coefficient of
inflation\textit{\ }$\mathit{k=1,2,..}$\textit{.}) can be written as 
\begin{equation}
\left\vert k\mathcal{P}\cap \mathbf{Z}^{n}\right\vert =\mathfrak{P}%
(k,n)=a_{n}(\mathcal{P})k^{n}+a_{n-1}(\mathcal{P})k^{n-1}+\cdot \cdot \cdot
+a_{0}(\mathcal{P})  \tag{18}
\end{equation}%
with coefficients $a_{0},...,a_{n}$ being specific for a given type of
polytope $\mathcal{P}$. In the case of Veneziano amplitude the polynomial $%
p(k,n)$ counts number of points inside the $n-$dimensional inflated simplex
(with inflation coefficient $k=1,2,...$). Irrespective to the polytope type,
it is known [38] that $a_{0}=1$ and $a_{n}=Vol\mathcal{P},$ where $Vol%
\mathcal{P}$ is the $\mathit{Euclidean}$ volume of the polytope. These facts
can be easily checked directly for $p(k,n).$ To calculate the remaining
coefficients of such polynomial explicitly for arbitrary convex rational
polytope $\mathcal{P}$ is a difficult task in general. Such a task was
accomplished only recently in [39]. The authors of [39] recognized that in
order to obtain the remaining coefficients, it is useful to calculate the
generating function for the Ehrhart polynomial. Long before the results of
[39] were published, it was known [3,27], that the generating function for
the Ehrhart polynomial of $\mathcal{P}$ can be written in the following
universal form 
\begin{equation}
\mathcal{F}(\mathcal{P},x)=\sum\limits_{k=0}^{\infty }\mathfrak{P}(k,n)x^{k}=%
\frac{h_{0}(P)+h_{1}(P)x+\cdot \cdot \cdot +h_{n}(P)x^{n}}{(1-x)^{n+1}}. 
\tag{19}
\end{equation}%
The above result \ leading to possible generalizations/extensions of the
Veneziano amplitudes does make physical sence as we shall demonstrate
momentarily. Additional details are also presented in Section 5.

The fact that the combinatorial factor $p(k,n)$ in Eq.(9) can be formally
written in several equivalent ways has physical significance. For instance,
in particle physics literature, e.g. see [2], the \textsl{third} option is
commonly used. Let us recall how this happens. One is looking for an
expansion of the factor $(1-x)^{-\alpha (t)-1}$ under the integral of beta
function, e.g. see Eq.(2). \ Looking at Eq.(19) one realizes that the Regge
variable $\alpha (t)$ plays the role of dimensionality of \textbf{Z-}
lattice. Hence, in view of Eq.(8), we have to identify it with $n$ $($or $k,$
in case if Eq.(8) is used) in the \textsl{second} option provided by Eq.(9).
This is not the way such an identification is done in physics literature
where, in fact, the \textsl{third} option\ in Eq.(9) is used with $k=\alpha
(t)$ \ being effectively the inflation factor while $n$ being\ effectively
the dimensionality of the lattice\footnote{%
We have to warn our readers that nowhere in physics literature such
combinatorial terminology is used to our knowledge.}. A quick look at
Eq.s(10) and (19) shows that under such circumstances the generating
function for the Ehrhart polynomial and that for the Veneziano amplitude are
formally \textsl{not} the same. In the first case one is dealing with
lattices of \textsl{fixed} dimensionality and is considering summation over
various inflation factors at the same time. In the second case (used in
physics literature [2]) one is dealing with the \textsl{fixed}\textit{\
inflation factor} $n=\alpha (t)$ while summing over lattices of different
dimensionalities. Nevertheless, such arguments are superficial in view of \
Eq.s(8) and (19) above. Using these equations it is clear that \textsl{%
mathematically correct} agreement between Eq.s(10) and (19) can be reached
only if one is using identification: $\mathfrak{P}(k,n)=p(k,n),$ with the 
\textsl{second} option given by Eq.(7) selected. By doing so no changes in
the pole locations for the Veneziano amplitude occur. Moreover, for a given
pole the second and the third option in Eq.(9) produce exactly the same
contributions into the residue thus making them physically
indistinguishable. The interpretation of the Veneziano amplitude as the
Laplace transform of the Ehrhart polynomial generating function provides a
very compelling reason for development of the alternative string-theoretic
formalism. In addition, it allows us to think about possible generalizations
of the Veneziano amplitude using generating functions for the Ehrhart
polynomials for polytopes \textsl{other} than the $n-$dimensional inflated
simplex used for the Veneziano amplitudes. As it is demonstrated by Stanley
[3,27], Eq.(19) has a group invariant meaning as the Poincare$^{\prime }$
polynomial for the so called Stanley-Reisner polynomial ring.\footnote{%
In Section 5 we provide some additional details on this topic.}. This fact
alone makes generalization of the Veneziano amplitudes mathematically
plausible. From the same reference one can find connections of these results
with toric varieties. \ In view of \ Ref.[14], this observation is
sufficient for restoration of physical models reproducing the Veneziano and
Veneziano-like generalized amplitudes. Thus, in the rest of this paper we
shall discuss some approaches to the design of these models.

Generalization of the Veneziano amplitudes is justified not only
mathematically. It is also needed physically as explained earlier in
Subsection 1.2. The information on Ehrhart polynomials just provided is
sufficient for this purpose as we would like to explain now.

\subsection{The generalized Veneziano amplitudes and mirror symmetry}

\bigskip

As we have explained already in Subsection 1.2., according to the Regge
theory [16,17], for each parent trajectory there should be a countable
infinity of daughter trajectories-all lying below the parent on the C-F
plot. \ In his original paper [1], page 195, Veneziano took this fact into
account and said explicitly that his amplitude is \textsl{not} uniquely
defined. Following both the original work by Veneziano and Ref.[15], p.100,
we notice that beta function in Eq.(2) given by $B(-\alpha (s),-\alpha (t))$
(which is effectively the unsymmetrized Veneziano amplitude) can be replaced
by $B(m-\alpha (s),n-\alpha (t))$ for any integers $m,n\geq 0$. To comply
with the Regge theory one should use \textsl{any} linear combination of beta
functions just described unless some additional assumptions are made. To our
knowledge, the fact that the Veneziano amplitude is not uniquely defined
regrettably is not mentioned in any of the existing modern string theory
literature. Hence, if the alternative (to ours) formulations of
string-theoretic models may finally produce some mathematically meanigful
results, these formulations still will be confronted with explanation of the
experimental fact that in nature only \textsl{finite} number of daughter
trajectories is observed for each parent trajectory. \ If one accepts the
viewpoint of this paper, such experimental fact can be explained quite
naturally with help of mirror symmetry arguments. It should be noted,
however, that our use of mirror symmetry differs drastically from that
currently in use [40,41]. Nevertheless, the initial observations used in the
present case do coincide with those used in more popular mirror symmetry
treatments [41] since in our case they are also based on the work by
Batyrev, Ref.[22]. In turn, \ Batyrev's results to some extent have been
influenced by the result of Hibi [24] to be used in our work as well.

Following these authors we would like to discuss properties of reflexive
(polar (or dual)) polytopes. It is useful to notice at this point that the
concept of the dual (polar) polytope was in use in solid state physics
literature [42] for quite some time. Indeed both direct and reciprocal
(dual) lattices are being used rutinely in calculations of physical
properties of crystalline solids. The requirement that physical observables
should remain the same irrespective to what lattice is used in calculations
is completely natural. The same, evidently, should be true in the mirror
symmetry calculations used in high energy physics. This is the \textsl{%
physicall essence} of mirror symmetry. \ In the paper by Greene and Plesser
[43], p.26, one finds the following statement : "Thus, we have demonstrated
that two topologically distinct \ Calabi-Yau manifolds $M$ and $\ M^{\prime }
$ give raise to \textsl{the same} conformal field theory. Furthermore,
although our argument has been based only at one point in the respective
moduli spaces $\mathcal{M}_{M}$ and $\mathcal{M}_{M^{\prime }}$ of $M$ amd $%
M^{\prime }($namely the point which has a minimal model interpretation and
hence respects the symmetries by which we have orbifolded) the results
necessarily extends to all of $\mathcal{M}_{M}$ and $\mathcal{M}_{M^{\prime
}}".$

We would like to explan these statements now using more commonly known
terminology. For this purpose \ we begin with the following

\ 

\textbf{Definition 2.1}. A subset of \ $\mathbf{R}^{d}$ is considered to be
a \textit{polytope (or polyhedron)} $\mathcal{P}$ if there is a $r\times d$
matrix $\mathbf{M}$\textbf{\ (}with\textbf{\ }$r\leq d)$ and a vector $%
\mathbf{b}\in \mathbf{R}^{d}$ such that $\mathcal{P}=\{\mathbf{x}\in \mathbf{%
R}^{d}\mid \mathbf{Mx}\leq \mathbf{b}\}.$ Provided that the Euclidean $d$%
-dimensional scalar product is given by $<\mathbf{x}\cdot \mathbf{y}%
>=\sum\limits_{i=1}^{d}x_{i}y_{i}$ , a \textit{rational (}respectively%
\textit{, integral) }polytope\textit{\ (or polyhedron)} $\mathcal{P}$ is
defined by the set 
\begin{equation}
\mathcal{P}=\{\mathbf{x}\in \mathbf{R}^{d}\mid <\mathbf{a}_{i}\cdot \mathbf{x%
}>\leq \beta _{i}\text{ , }i=1,...,r\},  \tag{20}
\end{equation}%
where $\mathbf{a}_{i}\in \mathbf{Q}^{d\frac{1}{\left( 1-t\right) ^{k+1}}}$
and $\beta _{i}\in \mathbf{Q}$ for $i=1,...,r$ (respectively $\mathbf{a}%
_{i}\in \mathbf{Z}^{d}$ and $\beta _{i}\in \mathbf{Z}$ for $i=1,...,r.).$

\bigskip

Next, we need yet another definition

\bigskip

\textbf{Definition 2.2. }For any convex polytope $\mathcal{P}$ the dual
polytope $\mathcal{P}^{\ast }$ is defined by 
\begin{equation}
\mathcal{P}^{\ast }=\{\mathbf{x}\in \left( R^{d}\right) ^{\ast }\mid
\left\langle \mathbf{a}\cdot \mathbf{x}\right\rangle \leq 1,\text{ }\mathbf{a%
}\in \mathcal{P}\}.  \tag{21}
\end{equation}

\ 

\ Although in algebraic geometry of toric varieties the inequality $%
\left\langle \mathbf{a}\cdot \mathbf{x}\right\rangle \leq 1$ is sometimes
replaced by $\left\langle \mathbf{a}\cdot \mathbf{x}\right\rangle \geq -1$
[38] we shall use the definition just stated to be in accord with Hibi [24].
According to this reference, if $\mathcal{P}$ is rational, then $\mathcal{P}%
^{\ast }$ is also rational. However, $\mathcal{P}^{\ast }$ is not
necessarily integral even if $\mathcal{P}$ is integral. This result is of
profound importance since the result, Eq.(19), is valid for the integral
polytopes only. The question arises : under what conditions is the dual
polytope $\mathcal{P}$ integral ? The answer is given by the following

\bigskip\ 

\textbf{Theorem 2.3.(}Hibi\textbf{\ [}24\textbf{]) }The dual polytope $%
\mathcal{P}^{\ast }$ is integral if and only if 
\begin{equation}
\mathcal{F}(\mathcal{P},x^{-1})=(-1)^{d+1}x\mathcal{F}(\mathcal{P},x) 
\tag{22}
\end{equation}%
where the generating function $\mathcal{F}(\mathcal{P},x)$ is defined in
Eq.(19).

\ 

By combining Eq.s(10) and (19) \ we obtain the following result for the
standard Veneziano amplitude%
\begin{equation}
\mathcal{F}(\mathcal{P},x)=\left( \frac{1}{1-x}\right) ^{d+1}.  \tag{23}
\end{equation}%
Using this expression in Eq.(22) produces:%
\begin{equation}
\mathcal{F}(\mathcal{P},x^{-1})=\frac{\left( -1\right) ^{d+1}}{\left(
1-x\right) ^{d+1}}x^{d+1}=(-1)^{d+1}x^{d+1}\mathcal{F}(\mathcal{P},x). 
\tag{24}
\end{equation}%
This result idicates that scattering processes described by the standard
Veneziano amplitudes \textsl{do not} involve any mirror symmetry since, as
it is well known [22,23] in order for such a symmetry to take place the dual
polytope $\mathcal{P}^{\ast }$ must be integral. In such a case both $%
\mathcal{P}$ and $\mathcal{P}^{\ast }$ \ encode (define) \ the projective
toric varieties $X_{P}$ and $X_{P^{\prime }}$ which are mirrors of each
other and are of Fano-type [22,23,45,46]. The question arises: can these
amplitudes be modified with help of Eq.(19) so that the presence of mirror
symmetry can be checked in nature? To answer this question, let us assume
that, indeed, Eq.(19) can be used for such a modification. In this case we
must require for the generating function $\mathcal{F}(\mathcal{P},x)$ in
Eq.(19) to obey Eq.(22). Direct check of such an assumption leads to the
desired result provided that $h_{n-i}=h_{i\text{ }}$in Eq$.(19).$
Fortunately, this \textsl{is} the case in view of the fact that these are
the famous Dehn- Sommerville equations, Ref.[38], p.16. Hence, at this stage
of our discussion, it looks like generalization of the Veneziano amplitudes
which takes into account mirror symmetry is possible from the mathematical
standpoint. Unfortunately, in physics correctness of mathematical arguments
is not sufficient for such generalization since experimental data may or may
not support such rigorous mathematics. To check the correctness of our
assupmtions (at least to a some extent) we would like to discuss now some
known in literature results on pion-pion \ $(\pi \pi )$ scattering
described, for example, in Refs.[25,46] from the point of view of results we
just obtained. By doing so we shall provide the evidence that: a) mirror
symmetry does exist in nature (wether or not its validity is nature's law or
just a curiocity remains to be further checked by analysing the available
experimental data) and, that b) use of mirror symmetry arguments permits us
to eliminate the countable infinity of daughter trajectories allowed by the
traditional Regge theory in favour of just several observed experimentally.

Experimentally it is known that, below the threshold, that is below the
collision energies producing more outgoing particles than incoming, the
unsymmetrized amplitude $A(s,t)$ for $\pi \pi $ scattering can be written as 
\begin{equation}
A(s,t)=-g^{2}\frac{\Gamma (1-\alpha (s))\Gamma (1-\alpha (t))}{\Gamma
(1-\alpha (s)-\alpha (t))}=-g^{2}(1-\alpha (s)-\alpha (t))B(1-\alpha
(s),1-\alpha (t)).  \tag{25}
\end{equation}%
This result should be understood as follows. Consider the "weighted" (still
unsymmetrized) Veneziano amplitude of the type%
\begin{equation}
A(s,t)=\dint\limits_{0}^{1}dxx^{-\alpha (s)-1}(1-x)^{-\alpha (t)-1}g(x,s,t) 
\tag{26}
\end{equation}%
where the weight function $g(x,s,t)$ is given by 
\begin{equation}
g(x,s,t)=\frac{1}{2}g^{2}[(1-x)\alpha (s)+x\alpha (t)].  \tag{27}
\end{equation}%
Upon integration, one recovers Eq.(25). The same result can be achieved if ,
instead one uses the weight function of the type%
\begin{equation}
g(x,s,t)=g^{2}x\alpha (t).  \tag{28}
\end{equation}%
In early treatments of the dual resonance models (all developed around the
Veneziano amplitude) [46] fitting to experimental data was achieved with
some ad hoc prescriptions for the weight function $g(x,s,t),e.g$ like those
given by Eq.s (27) and (28)$.$In the case of $\pi \pi $ scattering such an
ad hoc reasoning can be replaced by the requirements of mirror symmetry.
Indeed, consider a special case of Eq.(19): n=2. For such a case we obtain, 
\begin{equation}
\mathcal{F}(\mathcal{P},x)=\sum\limits_{k=0}^{\infty }\mathfrak{P}(k,n)x^{k}=%
\frac{h_{0}(P)+h_{1}(P)x}{(1-x)^{1+1}}  \tag{29}
\end{equation}%
so that Eq.(22) holds indicating mirror symmetry. At this point, in view of
Eq.s (26)-(29), one may notice that, actually, for this symmetry to take
place in real world, one should replace the amplitude given by Eq.(25) by
the following combination 
\begin{eqnarray}
A(s,t) &=&-g^{2}\frac{\Gamma (1-\alpha (s))\Gamma (1-\alpha (t))}{\Gamma
(1-\alpha (s)-\alpha (t))}+g^{2}\frac{\Gamma (-\alpha (s))\Gamma (-\alpha
(t))}{\Gamma (-\alpha (s)-\alpha (t))}  \notag \\
&=&-\hat{g}^{2}B(1-\alpha (s),1-\alpha (t))+g^{2}B(-\alpha (s),-\alpha (t)).
\TCItag{30}
\end{eqnarray}%
Such a combination produces first two terms (with correct signs) of the
infinite series as proposed by Mandelstam, Eq.(15) of Ref.[47]. The
comparison with experiment displayed in Fig.6.2(a) of Ref.[46], p.321, is
quite satisfactory producing one parent and one daughter Regge trajectories.
These are also displayed in Ref.[15], p. 41, for the "rho family" of
resonances. Thus, at least in the case of $\pi \pi $ scattering, one can
claim that mirror symmetry consideration provides a plausible explanation of
the observable data. One hopes, that the case just considered is typical so
that mirror symmetry does play a role in Nature.

The rest of this paper is devoted to the reconstruction of physical models
reproducing Veneziano and extended Veneziano amplitudes based on
mathematical results discussed in these two sections. Additional details can
be found in Refs.[10,11,34,48].

\section{\protect\bigskip\ Motivating examples}

To facilitate our readers understanding, we would like to illustrate general
principles using simple examples. We begin by considering a finite geometric
progression of the type 
\begin{align}
\mathcal{F(}c,m)& =\sum\limits_{l=-m}^{m}\exp \{cl\}=\exp
\{-cm\}\sum\limits_{l=0}^{\infty }\exp \{cl\}+\exp
\{cm\}\sum\limits_{l=-\infty }^{0}\exp \{cl\}  \notag \\
& =\exp \{-cm\}\frac{1}{1-\exp \{c\}}+\exp \{cm\}\frac{1}{1-\exp \{-c\}} 
\notag \\
& =\exp \{-cm\}\left[ \frac{\exp \{c(2m+1)\}-1}{\exp \{c\}-1}\right] . 
\tag{31}
\end{align}%
The reason for displaying the intermediate steps will be explained shortly
below. First, however, we would like to consider the limit : $c\rightarrow
0^{+}$ of $\mathcal{F(}c,m)$. It is given by $\mathcal{F(}0,m)=2m+1$. The
number $2m+1$ equals to the number of integer points in the segment $[-m,m]$ 
\textit{including} \textit{boundary} points. It is convenient to rewrite the
above result in terms of $x=\exp \{c\}$ so that we shall write formally $%
\mathcal{F(}x,m)$ instead of $\mathcal{F(}c,m)$ from now on. Using such
notation, consider a related function 
\begin{equation}
\mathcal{\bar{F}(}x,m)=(-1)\mathcal{F(}\frac{1}{x},-m).  \tag{32}
\end{equation}%
This type of relation (the \textit{Ehrhart-Macdonald reciprocity law}) is
characteristic for the Ehrhart polynomial for rational polytopes discussed
earlier. In the present case we obtain, 
\begin{equation}
\mathcal{\bar{F}(}x,m)=(-1)\frac{x^{-(-2m+1)}-1}{x^{-1}-1}x^{m}.  \tag{33}
\end{equation}%
In the limit $x\rightarrow 1+0^{+}$ we obtain : $\mathcal{\bar{F}(}%
1,m)=2m-1. $ The number $2m-1$ is equal to the number of integer points
strictly \textit{inside} the segment $[-m,m].$ \ Both $\mathcal{F(}0,m)$ and 
$\mathcal{\bar{F}(}1,m)$ \ provide the simplest possible examples of the
Ehrhart polynomials if we identify $m$ with the inflation factor $k$.

These, seemingly trivial, results can be broadly generalized. First, we
replace $x$ by $\mathbf{x=}x_{1}\cdot \cdot \cdot x_{d}$, next we replace
the summation sign in the left hand side of Eq.(31) by the multiple
summation, etc. Thus obtained function $\mathcal{F(}\mathbf{x},m)$ in the
limit $x_{i}\rightarrow 1+0^{+},$ $i=1-d,$ produces the anticipated result $:%
\mathcal{F(}\mathbf{1},m)=(2m+1)^{d}$ . It describes the number of points
inside and at the faces of $\ $a $d-$ dimensional cube in the Euclidean
space $\mathbf{R}^{d}$. Accordingly, for the number of points strictly
inside the cube we obtain : $\mathcal{\bar{F}(}\mathbf{1},m)=(2m-1)^{d}.$

Let \textit{Vert}$\mathcal{P}$ denote the vertex set of the rational
polytope. In the case considered thus far it is the $d-$dimensional cube.
Let $\{u_{1}^{v},...,u_{d}^{v}\}$ denote the orthogonal basis (not
necessarily of unit length) made of the highest weight vectors of the
Weyl-Coxeter reflection group $B_{d}$ appropriate for cubic symmetry [11].
These vectors are oriented along the positive semi axes with respect to the
center of symmetry of the cube. When parallel translated to the edges ending
at particular hypercube vertex \textbf{v}, they can point either in or out
of this vertex. Then, the $d$-dimensional version of Eq.(31) can be
rewritten in notations just introduced as follows 
\begin{equation}
\sum\limits_{\mathbf{x}\in \mathcal{P\cap }\mathbf{Z}^{d}}\exp \{<\mathbf{c}%
\cdot \mathbf{x}>\}=\sum\limits_{\mathbf{v}\in Vert\mathcal{P}}\exp \{<%
\mathbf{c}\cdot \mathbf{v}>\}\left[ \prod\limits_{i=1}^{d}(1-\exp
\{-c_{i}u_{i}^{v}\})\right] ^{-1}.  \tag{34}
\end{equation}%
The correctness of this equation can be readily checked by considering
special cases of a segment, square, cube, etc. The result, Eq.(34), obtained
for the polytope of cubic symmetry can be extended to the arbitrary convex
centrally symmetric polytope. Details can be found in Ref.[49]. Moreover,
the requirement of central symmetry can be relaxed to the requirement of the
convexity of \ $\mathcal{P}$ only. In such general form the relation given
by Eq.(34) was obtained by Brion [50]. It is of central importance for the
purposes of this work: the limiting procedure $c\rightarrow 0^{+}$ produces
the number of points inside (and at the boundaries) of the polyhedron $%
\mathcal{P}$ in the l.h.s. of Eq.(34) and, if the polyhedron is rational and
inflated, this procedure produces the Ehrhart polynomial. Actual
computations are done with help of the r.h.s. of Eq.(34) as will be
demonstrated below.

\section{\protect\bigskip\ The Duistermaat-Heckman formula and the
Khovanskii-Pukhlikov correspondence}

Since the description of the Duistermaat-Heckman \ (D-H) formula can be
found in many places, we would like to be brief \ in discussing it now in
connection with earlier obtained results. Let $M\equiv M^{2n}$ be a compact
symplectic manifold equipped with the moment map $\Phi :M\rightarrow \mathbf{%
R}$ and the (Liouville) volume form $dV=\left( \frac{1}{2\pi }\right) ^{n}%
\frac{1}{n!}$ $\Omega ^{n}.$\ According to the Darboux theorem, locally $%
\Omega =\sum\nolimits_{l=1}^{n}$\ \ $dq_{l}$\ $\wedge dp_{l}$\ . We expect
that such a manifold has isolated fixed points $p$ belonging to the fixed
point set $\mathcal{V}$ associated with the isotropy subgroup of the group $%
G $ acting on $M$. Then, in its most general form, the D-H formula can be
written as [14,26,51] 
\begin{equation}
\int\limits_{M}dVe^{\Phi }=\sum\limits_{p\in \mathcal{V}}\frac{e^{\Phi (p)}}{%
\prod\nolimits_{j}a_{j,p}},  \tag{35}
\end{equation}%
where $a_{1,p},...,a_{n,p}$ are the weights of the linearized action of $G$
on $T_{p}M$. Using Morse theory, Atiyah [52] and others, e.g. see Ref.[54]
for additional references, have demonstrated that it is sufficient to keep
terms up to quadratic in the expansion of $\Phi $ around given $p$. In such
a case the moment map can be associated with the Hamiltonian for the finite
set of harmonic oscillators. In the properly chosen system of units the
coefficients $a_{1,p},...,a_{n,p}$ are just \textquotedblright
masses\textquotedblright\ $m_{i}$ \ of the individual oscillators. Unlike
truly physical masses, some of $m_{i}^{\prime }s$ can be negative.

Based on the information just provided, we would like to be more specific
now. To this purpose, following Vergne [53] and Brion [50], we would like to
consider the D-H integral of the form 
\begin{equation}
I(k\text{ };\text{ }y_{1},y_{2})=\int\nolimits_{k\Delta }dx_{1}dx_{2}\exp
\{-(y_{1}x_{1}+y_{2}x_{2})\},  \tag{36}
\end{equation}%
where $k\Delta $ is the standard dilated simplex with \ dilation coefficient 
$k\footnote{%
Our choice of the simplex as the domain of integration is caused by our
earlier made observation [10] that the deformation retract of the Fermat
(hyper)surface (on which the Veneziano amplitude lives ) is just the
standard simplex. Since such Fermat surface is a complex K\"{a}hler-Hodge
type manifold and since all K\"{a}hler manifolds are symplectic [26,54], our
choice makes sense.}$. Calculation of \ this integral can be done exactly
with the result: 
\begin{equation}
I(k\text{ };\text{ }y_{1},y_{2})=\frac{1}{y_{1}y_{2}}+\frac{e^{-ky_{1}}}{%
y_{1}(y_{1}-y_{2})}+\frac{e^{-ky_{2}}}{y_{2}(y_{2}-y_{1})}  \tag{37}
\end{equation}%
consistent with Eq.(35). In the limit: $y_{1},y_{2}\rightarrow 0$ some
calculation produces the anticipated result $:Volk\Delta =k^{2}/2!$ \ for
the \textit{Euclidean} volume of the dilated simplex. Next, to make a
connection with the previous section, in particular, with Eq.(34), consider
the following sum 
\begin{align}
S(k\text{ ; }y_{1}\text{,}y_{2})& =\sum\limits_{\left( l_{1},l_{2}\right)
\in k\Delta }\exp \{-(y_{1}l_{1}+y_{2}l_{2})\}  \notag \\
& =\frac{1}{1-e^{-y_{1}}}\frac{1}{1-e^{-y_{2}}}+\frac{1}{1-e^{y_{1}}}\frac{%
e^{-ky_{1}}}{1-e^{y_{1}-y_{2}}}+\frac{1}{1-e^{y_{2}}}\frac{e^{-ky_{2}}}{%
1-e^{y_{2}-y_{1}}}  \tag{38}
\end{align}%
related to the D-H integral, Eq.s(36,37). Its calculation will be explained
momentarily. In spite of the connection with the D-H integral, the limiting
procedure: $y_{1},y_{2}\rightarrow 0$ in the last case is much harder to
perform. It is facilitated by use of the following expansion 
\begin{equation}
\frac{1}{1-e^{-s}}=\frac{1}{s}+\frac{1}{2}+\frac{s}{12}+O(s^{2}).  \tag{39}
\end{equation}%
Rather lenghty calculations produce the anticipated result : $S(k$ ; $0$,$%
0)=k^{2}/2!+3k/2+1\equiv \left\vert k\Delta \cap Z^{2}\right\vert \equiv 
\mathfrak{P}(k,2)$ for the Ehrhart polynomial. Since generalization of the
obtained results to simplicies of higher dimensions is straightforward, the
relevance of these results to the Veneziano amplitude should be evident. To
make it more explicit we have to make several steps still. First, we would
like to explain how the result, Eq.(38), was obtained. By doing so we shall
gain some additional physical information. Second, we would like to explain
in some detail the connection between the integral, Eq.(37), and the sum,
Eq.(38). Such a connection is made with help of the Khovanskii-Pukhlikov
correspondence.

We begin with calculations of the sum, Eq.(38). To do this we need a
definition of the convex rational polyhedral cone $\sigma .$ It is given by 
\begin{equation}
\sigma =\mathbf{Z}_{\geq 0}a_{1\text{\ }}+\cdot \cdot \cdot +\mathbf{Z}%
_{\geq 0}a_{d\text{\ }},  \tag{40}
\end{equation}%
where the set $a_{1},...,a_{d}$ \ forms a basis (not nesessarily orthogonal)
of the $d$-dimensional vector space $V,$ while $\mathbf{Z}_{\geq 0}$ are non
negative integers. It is known that all combinatorial information about the
polytope $\mathcal{P}$ is encoded in the $\mathit{complete}$ fan made of
cones whose apexes all having the same origin in common. Details can be
found in literature [26,30]. At the same time, the vertices of $\mathcal{P}$
are also the apexes of the respective cones. Following Brion[50], this fact
allows us to write the l.h.s. of Eq.(34) as 
\begin{equation}
f(\mathcal{P},x)=\sum\limits_{\mathbf{m}\in \mathcal{P}\cap \mathbf{Z}^{d}}%
\mathbf{x}^{\mathbf{m}}=\sum\limits_{\sigma \in Vert\mathcal{P}}\mathbf{x}^{%
\mathbf{\sigma }}  \tag{41}
\end{equation}%
so that for the \textit{dilated} polytope the above statement reads as
follows [50,55]: 
\begin{equation}
f(k\mathcal{P},x)=\sum\limits_{\mathbf{m}\in k\mathcal{P}\cap \mathbf{Z}^{d}}%
\mathbf{x}^{\mathbf{m}}=\sum\limits_{i=1}^{n}\mathbf{x}^{\mathbf{kv}%
_{i}}\sum\limits_{\sigma _{i}}\mathbf{x}^{\mathbf{\sigma }_{i}}.  \tag{42}
\end{equation}%
In the last formula the summation is taking place over all vertices whose
location is given by the vectors from the set \{$\mathbf{v}_{1},...,\mathbf{v%
}_{n}\}.$ This means that in actual calculations one can first calculate the
contributions coming from the cones $\sigma _{i}$ of the undilated
(original) polytope $\mathcal{P}$ and only then one can use the last
equation in order to get the result for the dilated polytope.

Let us apply these general results to our specific problem of computation of 
$S(k$ ; $y_{1}$,$y_{2})$ in Eq.(38). We have our simplex with vertices in
x-y plane given by the vector set \{ \textbf{v}$_{1}$=$(0,0)$, \textbf{v}$%
_{2}$=(1,0), \textbf{v}$_{3}$=(0,1)\}, where we have written the x
coordinate first. In this case we have 3 cones: $\sigma _{1}=l_{2}\mathbf{v}%
_{2}+l_{3}\mathbf{v}_{3}$ , $\sigma _{2}=\mathbf{v}_{2}+l_{1}(-$\textbf{v}$%
_{2})+l_{2}($\textbf{v}$_{3}-$\textbf{v}$_{2});\sigma _{3}=$\textbf{v}$%
_{3}+l_{3}($\textbf{v}$_{2}-$\textbf{v}$_{3})+l_{1}(-\mathbf{v}_{3});$\{$%
l_{1}$, $l_{2}$ , $l_{3}$ \}$\in \mathbf{Z}_{\geq 0}$ . In writing these
expressions for the cones we have taken into account that, according to
Brion, when making calculations the apex of each cone should be chosen as
the origin of the coordinate system. Calculation of contributions to the
generating function coming from $\sigma _{1}$ is the most straightforward.
Indeed, in this case we have $\mathbf{x}=x_{1}x_{2}=e^{-y_{1}}e^{-y_{2}}.$
Now, the symbol $\mathbf{x}^{\mathbf{\sigma }}$ in Eq.(41) should be
understood as follows. Since $\sigma _{i}$ , $i=1-3$, is actually a vector,
it has components, like those for \textbf{v}$_{1},$ etc$.$ We shall write
therefore $\mathbf{x}^{\mathbf{\sigma }}=x_{1}^{\sigma (1)}\cdot \cdot \cdot
x_{d}^{\sigma (d)}$ where $\sigma (i)$ is the i-th component of such a
vector. Under these conditions calculation of the contributions from the
first cone with the apex located at (0,0) is completely straightforward and
is given by 
\begin{equation}
\sum\limits_{\left( l_{2},l_{3}\right) \in
Z_{+}^{2}}x_{1}^{l_{2}}x_{2}^{l_{3}}=\frac{1}{1-e_{{}}^{-y_{1}}}\frac{1}{%
1-e^{-y_{2}}}.  \tag{43}
\end{equation}%
It is reduced to the computation of the infinite geometric progression. But
physically, the above result can be looked upon as a product of two
partition functions for two harmonic oscillators whose ground state energy
was discarded. By doing the rest of calculations in the way just described
we reobtain $S(k$ ; $y_{1}$,$y_{2})$ from Eq.(38) as required. This time,
however, we know that the obtained result is associated with the assembly of
harmonic oscillators of frequencies $\pm y_{1}$ and $\pm y_{2}$ and $\pm
(y_{1}-y_{2})$ whose ground state energy is properly adjusted. The
\textquotedblright frequencies\textquotedblright\ (or masses) of these
oscillators are coming from the Morse-theoretic considerations for the
moment maps associated with the critical points of symplectic manifolds as
explained \ in the paper by Atiyah [52]. These masses enter into the
\textquotedblright classical\textquotedblright\ D-H formula, Eq.s(36),(37).
It is just a classical partition function for a system of such described
harmonic oscillators living in the phase space containing critical points.
The D-H classical partition function, Eq.(37), has its quantum analog,
Eq.(38), just described. The ground state for such a quantum system is
degenerate with the degeneracy being described by the Ehrhart polynomial $%
\mathfrak{P}(k,2).$ Such a conclusion is in formal accord with results of
Vergne [14].

Since (by definition) the coefficient of dilation \textit{k=1,2,... , 
\textsl{there is no dynamical system (and its quanum analog) for k=0}. 
\textsl{But this condition is the} \textsl{condition for existence of the
tachyon pole in the Veneziano amplitude, Eq.(2)}. \textsl{Hence, in view of
the results just described this pole should be} \textsl{considered as
unphysical and discarded.} }Such arguments are independent of the analysis
made in Ref.[10] where the unphysical tachyons are removed with help of the
properly adjusted phase factors. Clearly, such factors can be reinstated in
the present case as well since their existence is caused by the requirements
of the torus action invariance of the Veneziano-like amplitudes as explained
in [10,26]. Hence, their presence is consistent with results just presented.

Now we are ready to discuss the Khovanskii-Pukhlikov correspondence. It can
be understood based on the following generic example taken from Ref.[56]. We
would like to compare the integral 
\begin{equation*}
I(z)=\int\limits_{s}^{t}dxe^{zx}=\frac{e^{tz}}{z}-\frac{e^{sz}}{z}\text{ \
with the sum }S(z)=\text{ }\sum\limits_{k=s}^{t}e^{kz}=\frac{e^{tz}}{1-e^{-z}%
}+\frac{e^{sz}}{1-e^{z}}.
\end{equation*}%
To do so, following Refs[56-58] we introduce the Todd operator (transform)
via 
\begin{equation}
Td(z)=\frac{z}{1-e^{-z}}.  \tag{44}
\end{equation}%
Then, it can be demonstrated that 
\begin{equation}
Td(\frac{\partial }{\partial h_{1}})Td(\frac{\partial }{\partial h_{2}}%
)(\int\nolimits_{s-h_{1}}^{t+h_{2}}e^{zx}dx)\mid
_{h_{1}=h_{1}=0}=\sum\limits_{k=s}^{t}e^{kz}.  \tag{45}
\end{equation}%
This result can be now broadly generalized. Following Khovanskii and
Pukhlikov [57], we notice that 
\begin{equation}
Td(\frac{\partial }{\partial \mathbf{z}})\exp \left(
\sum\limits_{i=1}^{n}p_{i}z_{i}\right) =Td(p_{1},...,p_{n})\exp \left(
\sum\limits_{i=1}^{n}p_{i}z_{i}\right) .  \tag{46}
\end{equation}%
By applying this transform to 
\begin{equation}
i(x_{1},...,x_{k};\xi _{1},...,\xi _{k})=\frac{1}{\xi _{1}...\xi _{k}}\exp
(\sum\limits_{i=1}^{k}x_{i}\xi _{i})  \tag{47}
\end{equation}%
we obtain, 
\begin{equation}
s(x_{1},...,x_{k};\xi _{1},...,\xi _{k})=\frac{1}{\prod\limits_{i=1}^{k}(1-%
\exp (-\xi _{i}))}\exp (\sum\limits_{i=1}^{k}x_{i}\xi _{i}).  \tag{48}
\end{equation}%
This result should be compared now with the individual terms on the r.h.s.
of Eq.(34) on one hand and with the individual terms on the r.h.s of Eq.(35)
on another. Evidently, with help of the Todd transform the exact
\textquotedblright classical\textquotedblright\ results for the D-H integral
are transformed into the \textquotedblright quantum\textquotedblright\
results of the Brion's identity, Eq.(34), which is actually equivalent to
the Weyl character formula [48].

We would like to illustrate these general observations by comparing the D-H
result, Eq.(37), with the Weyl character formula result, Eq.(38). To this
purpose we need to use already known data for the cones $\sigma _{i}$ , $%
i=1-3,$ and the convention for the symbol \textbf{x}$^{\sigma }$. In
particular, \ for the first cone we have already $:\mathbf{x}^{\sigma
_{1}}=x_{1}^{l_{1}}x_{2}^{l_{2}}=\left[ \exp (l_{1}y_{1})\right] \cdot \left[
\exp (l_{2}y_{2})\right] \footnote{%
To obtain correct results we needed to change\ signs in front of $l_{1}$ and 
$l_{2}$ . The same should be done for other cones as well.}.$ Now we
assemble the contribution from the first vertex using Eq.(37). We obtain, $%
\left[ \exp (l_{1}y_{1})\right] \cdot \left[ \exp (l_{2}y_{2})\right]
/y_{1}y_{2}.$ Using the Todd transform we obtain, 
\begin{equation}
Td(\frac{\partial }{\partial l_{1}})Td(\frac{\partial }{\partial l_{2}})%
\frac{1}{y_{1}y_{2}}\left[ \exp (l_{1}y_{1})\right] \cdot \left[ \exp
(l_{2}y_{2})\right] \mid _{l_{1}=l_{2}=0}=\frac{1}{1-e^{-y_{1}}}\frac{1}{%
1-e^{-y_{2}}}.  \tag{49}
\end{equation}%
Analogously, for the second cone we obtain: $\mathbf{x}_{2}^{\sigma
}=e^{-}ky_{1}e^{-}l_{1}y_{1}e^{-}l_{2}(y_{1}-y_{2})$ \ so that use of the
Todd transform produces: 
\begin{equation}
Td(\frac{\partial }{\partial l_{1}})Td(\frac{\partial }{\partial l_{2}})%
\frac{1}{y_{1}\left( y_{1}-y_{2}\right) }%
e^{-ky_{1}}e^{-l_{1}y_{1}}e^{-l_{2}(y_{1}-y_{2})}\mid _{l_{1}=l_{2}=0}=\frac{%
1}{1-e^{y_{1}}}\frac{e^{-ky_{1}}}{1-e^{y_{1}-y_{2}}},  \tag{50}
\end{equation}%
etc.

The obtained results can now be broadly generalized. To this purpose we can
formally rewrite the partition function, Eq.(24), in the following symbolic
form 
\begin{equation}
I(k,\mathbf{f})=\int\limits_{k\Delta }d\mathbf{x}\exp \mathbf{(-f\cdot x)} 
\tag{51}
\end{equation}%
valid for any finite dimension $d$. Since we have performed all calculations
explicitly for two dimensional case, for the sake of space, we only provide
the idea behind such type of calculation\footnote{%
Mathematically inclined reader is encoraged to read paper by Brion and
Vergne, Ref.[59], where all missing details are scrupulously presented.}. In
particular, using Eq.(37) we can rewrite this integral formally as follows 
\begin{equation}
\int\limits_{k\Delta }d\mathbf{x}\exp \mathbf{(-f\cdot x)=}\sum\limits_{p}%
\frac{\exp (-\mathbf{f}\cdot \mathbf{x(}p\mathbf{)})}{\prod%
\limits_{i}^{d}h_{i}^{p}(\mathbf{f})}.  \tag{52}
\end{equation}%
Applying the Todd operator (transform) to both sides of this formal
expression and taking into account Eq.s(49),(50) (providing assurance that
such an operation indeed is legitimate and makes sense) we obtain, 
\begin{align}
\int\limits_{k\Delta }d\mathbf{x}\prod\limits_{i=1}^{d}\frac{x_{i}}{1-\exp
(-x_{i})}\exp \mathbf{(-f\cdot x)}& =\sum\limits_{\mathbf{v}\in Vert\mathcal{%
P}}\exp \{<\mathbf{f}\cdot \mathbf{v}>\}\left[ \prod\limits_{i=1}^{d}(1-\exp
\{-h_{i}^{v}(\mathbf{f})u_{i}^{v}\})\right] ^{-1}  \notag \\
& =\sum\limits_{\mathbf{x}\in \mathcal{P\cap }\mathbf{Z}^{d}}\exp \{<\mathbf{%
f}\cdot \mathbf{x}>\},  \tag{53}
\end{align}%
where the last line was written in view of Eq.(34). From here, in the limit
: $\mathbf{f}=0$ we reobtain $p(k,n)$ defined in Eq.(10). Thus, using
classical partition function, Eq.(51), (discussed in the form of Exercises
2.27 and 2.28 \ in the book, Ref.[58], by Guillemin) and applying to it the
Todd transform we recover the quantum mechanical partition function whose
ground state provides us with the combinatorial factor $p(k,n).$

\section{\protect\bigskip From analysis to synthesis}

\subsection{\protect\bigskip The Poincare$^{\prime }$ polynomial}

The results discussed earlier are obtained for some fixed dilation factor $k$%
. In view of \ Eq.(8), \ they can be \ rewritten in the form valid for any
dilation factor $k$. To this purpose it is convenient to rewrite Eq.(8) in
the following equivalent form: 
\begin{align}
\frac{1}{\det (1-Mt)}& =\frac{1}{(1-tz_{0})\cdot \cdot \cdot (1-tz_{k})}%
=(1+tz_{0}+\left( tz_{0}\right) ^{2}+...)\cdot \cdot \cdot (1+tz_{n}+\left(
tz_{n}\right) ^{2}+...)  \notag \\
& =\sum\limits_{n=0}^{\infty
}(\sum\limits_{k_{0}+...+k_{k}=n}z_{0}^{k_{0}}\cdot \cdot \cdot
z_{k}^{k_{k}})t^{n}\equiv \sum\limits_{n=0}^{\infty }tr(M_{n})t^{n}, 
\tag{54}
\end{align}%
where the linear map from $k+1$ dimensional vector space $V$ to $V$ is given
by $\ $matrix $M\in G\subset GL(V)$ whose eigenvalues are $z_{0},...,z_{k}.$
Using this observation several conclusions can be drawn. First, it should be
clear that 
\begin{equation}
\sum\limits_{k_{0}+...+k_{k}=n}z_{0}^{k_{0}}\cdot \cdot \cdot
z_{k}^{k_{k}}=\sum\limits_{\mathbf{m}\in n\Delta \cap \mathbf{Z}^{k+1}}%
\mathbf{x}^{\mathbf{m}}=tr(M_{n}).  \tag{55}
\end{equation}%
Second, following Stanley [3, 27] we would like to consider the algebra of
invariants of $G$. To this purpose we introduce a basis $\mathbf{x}=$\{ $%
x_{0},...,x_{k}\}$ of $V$ and the polynomial ring $R=\mathbf{C}$[$%
x_{0},...,x_{k}]$ so that if $f\in R$ , then $Mf(\mathbf{x})=f(M\mathbf{x})$%
. The algebra of invariant polynomials $R^{G}$ can be defined now as 
\begin{equation*}
R^{G}=\{f\in R:Mf(\mathbf{x})=f(M\mathbf{x})=f(\mathbf{x})\text{ \ }\forall
M\in G\}.
\end{equation*}%
These invariant polynomials can be explicitly constructed as averages over
the group $G$ according to prescription: 
\begin{equation}
Av_{G}f=\frac{1}{\left\vert G\right\vert }\sum\limits_{M\in G}Mf,  \tag{56}
\end{equation}%
with $\left\vert G\right\vert $ being the cardinality of $G$. Suppose now
that $f\in R^{G}$, then, evidently, $f\in R^{G}=Av_{G}f$ so that $%
Av_{G}^{2}f=Av_{G}f=f$ . Hence, the operator $Av_{G}$ is indepotent. Because
of this, its eigenvalues can be only 1 and 0. From here it follows that 
\begin{equation}
\dim f_{n}^{G}=\frac{1}{\left\vert G\right\vert }\sum\limits_{M\in
G}tr(M_{n}).  \tag{57}
\end{equation}%
Thus far our analysis was completely general. To obtain Eq.(9) we have to
put $z_{0}=...=z_{k}=1$ in Eq.(8). This time, however we can use the
obtained results in order to write the following expansion for the Poincar$%
e^{\prime }$ polynomial [3, 27] which for the appropriately chosen $G$ is
equivalent to Eq.(10): 
\begin{equation}
P(R^{G},t)=\sum\limits_{n=0}^{\infty }\frac{1}{\left\vert G\right\vert }%
\sum\limits_{M\in G}tr(M_{n})t^{n}=\sum\limits_{n=0}^{\infty }\dim
f_{n}^{G}t^{n}.  \tag{58}
\end{equation}%
Evidently, the Ehrhart polynomial $\mathfrak{P}(k,n)=\dim f_{n}^{G}.$ To
figure out the group $G$ in the present case is easy since, actually, \ the
group is trivial: $G=1.$This is so because the eigenvalues $z_{0}$, ..., $%
z_{k}$ of the matrix $M$ all are equal to 1. It should be clear, however,
that for some appropriately chosen group $G$ expansion (19) is also the
Poincare polynomial (for the Cohen -Macaulay polynomial algebra [3,27]).
This fact provides independent (of Refs. [10,11]) evidence that both the
Veneziano and Veneziano-like amplitudes are of topological origin.

\subsection{Connections with intersection theory}

We would like to strengthen this observation now. To this purpose, in view
of Eq.(35), and taking into account that for the symplectic 2-form $\Omega
=\sum\nolimits_{i=1}^{k}dx_{i}\wedge dy_{i}$ the $n$-th power is given by $%
\Omega ^{n}=\Omega \wedge \Omega \wedge \cdot \cdot \cdot \wedge \Omega $ =$%
dx_{1}\wedge dy_{1}\wedge \cdot \cdot \cdot dx_{n}\wedge dy_{n}$, it is
convenient to introduce the differential form 
\begin{equation}
\exp \Omega =1+\Omega +\frac{1}{2!}\Omega \wedge \Omega +\frac{1}{3!}\Omega
\wedge \Omega \wedge \Omega +\cdot \cdot \cdot \text{ \ \ \ .}  \tag{59}
\end{equation}%
By design, the expansion \ in the r.h.s. will have only $k$ terms. The form $%
\Omega $ is closed, i.e. $d\Omega =0$ (the Liouvolle theorem), but not
exact. In view of the expansion, Eq.(59), the D-H integral, Eq.(51) can be
rewritten as 
\begin{equation}
I(k,\mathbf{f})=\int\limits_{k\Delta }\exp \mathbf{(}\tilde{\Omega}\mathbf{),%
}  \tag{60}
\end{equation}%
where, following Atiyah and Bott [28], we have introduced the form $\tilde{%
\Omega}=\Omega -\mathbf{f\cdot x.}$ Doing so requires us to replace the
exterior derivative $d$ acting on $\Omega $ by $\tilde{d}=d+i(\mathbf{x})$ \
(where the operator $i(\mathbf{x})$ reduces the degree of the form by one)
with respect to which the form $\tilde{\Omega}$ is equivariantly closed, i.e.%
$\tilde{d}\tilde{\Omega}=0.$ More explicitly, we have $\tilde{d}\tilde{\Omega%
}=d\Omega +i(\mathbf{x})\Omega -\mathbf{f\cdot dx=}0$. Since $d\Omega =0,$
we obtain the equation for the moment map : $i(\mathbf{x})\Omega -\mathbf{%
f\cdot dx=}0$ [51,58]. If use of the operator $d$ on differential forms
leads to the notion of cohomology, use of the operator $\tilde{d}$ leads to
the notion of equivariant cohomology. Although details can be found in the
paper by Atiyah and Bott [28], more relaxed pedagogical exposition can be
found in the monograph by Guillemin and Sternberg [60]. To make further
progress, we would like to rewrite the two-form $\Omega $ in complex
notations [51]. To this purpose, we introduce $z_{j}=p_{j}+iq_{j}$ and its
complex conjugate. In terms of these variables $\Omega $ acquires the
following form $:\Omega =$ $\frac{i}{2}\sum\nolimits_{i=1}^{k}dz_{i}\wedge d%
\bar{z}_{i}.$ Next, recall [61] that for any K\"{a}hler manifold the
fundamental 2-form $\Omega $ can be written as $\Omega =\frac{i}{2}%
\sum\nolimits_{ij}h_{ij}(z)dz_{i}\wedge d\bar{z}_{j}$ \ provided that $%
h_{ij}(z)=\delta _{ij}+O(\left\vert z\right\vert ^{2})$. This means that in
fact all K\"{a}hler manifolds are symplectic [26,54]. On such K\"{a}hler
manifolds one can introduce the Chern cutrvature 2-form which (up to a
constant ) should look like $\Omega .$ \ It should belong to the first Chern
class [19]. This means that, at least formally, consistency reqiures us to
identify x$_{i}^{\prime }s$ entering the product $\mathbf{f\cdot x}$ in the
form $\tilde{\Omega}$ with the first Chern classes $c_{i}$, i.e. $\mathbf{%
f\cdot x\equiv }\sum\nolimits_{i=1}^{d}f_{i}c_{i}.$ This fact was proven
rigorously in the above mentioned paper by Atiyah and Bott [28]. Since in
the Introduction we already mentioned that the Veneziano amplitudes can be
formally associated with the period integrals for the Fermat (hyper)surfaces 
$\mathcal{F}$ and since such integrals can be interpreted as intersection
numbers between the cycles on $\mathcal{F}$ [13,28,61] (see also Ref.[58],
p.72) one can formally rewrite the \textit{precursor} to the Veneziano
amplitude [10] as 
\begin{equation}
I=\left( \frac{-\partial }{\partial f_{0}}\right) ^{r_{0}}\cdot \cdot \cdot
\left( \frac{-\partial }{\partial f_{d}}\right) ^{r_{d}}\int\limits_{\Delta
}\exp (\tilde{\Omega})\mid _{f_{i}=0\text{ }\forall i}=\int\limits_{\Delta }d%
\mathbf{x(}c_{0})^{r_{0}}\cdot \cdot \cdot (c_{d})^{r_{d}}  \tag{61}
\end{equation}%
provided that $r_{0}+\cdot \cdot \cdot +r_{d}=n$ \ in in view of Eq.(13).
Analytical continuation of such an integral (as in the case of usual beta
function) then will produce the Veneziano amplitudes. In such a language,
calculation of the Veneziano amplitudes using generating function, Eq.(60),
mathematically becomes almost equivalent to calculations of averages in the
Witten-Kontsevich model [31-33]\footnote{%
This fact is explained in more details in Ref.[34].}. In \ addition, as was
also noticed by Atiyah and Bott [28], the replacement of the exterior
derivative $d$ by $\tilde{d}=d+i(\mathbf{x})$ was inspired by earlier work
by Witten on supersymmetric formulation of quantum mechanics and Morse
theory [29]. Such an observation along with results of Ref.[60] allows us to
develop calculations of the Veneziano amplitudes using supersymmetric
formalism.

\subsection{Supersymmetry and the Lefshetz isomomorphism}

We begin with the following observations. Let $X$ be the complex Hermitian
manifold and let $\mathcal{E}^{p+q}(X)$ denote the complex -valued
differential forms (sections) of type $(p,q),p+q=r,$ living on $X$. The
Hodge decomposition insures that $\mathcal{E}^{r}(X)$=$\sum\nolimits_{p+q=r}%
\mathcal{E}^{p+q}(X).$ The Dolbeault operators $\partial $ and $\bar{\partial%
}$ act on $\mathcal{E}^{p+q}(X)$ according to the rule \ $\partial :\mathcal{%
E}^{p+q}(X)\rightarrow \mathcal{E}^{p+1,q}(X)$ and $\bar{\partial}:\mathcal{E%
}^{p+q}(X)\rightarrow \mathcal{E}^{p,q+1}(X)$ , so that the exterior
derivative operator is defined as $d=\partial +\bar{\partial}$. Let now $%
\varphi _{p}$,$\psi _{p}\in \mathcal{E}^{p}$. By analogy with traditional
quantum mechanics we define (using Dirac's notations) the inner product 
\begin{equation}
<\varphi _{p}\mid \psi _{p}>=\int\limits_{M}\varphi _{p}\wedge \ast \bar{\psi%
}_{p},  \tag{62}
\end{equation}%
where the bar means the complex conjugation and the star $\ast $ means the
usual Hodge conjugation. Use of such a product is motivated by the fact that
the period integrals, e.g. those for the Veneziano-like amplitudes, and,
hence, those given by Eq.(61), are expressible through such inner products
[61]. Fortunately, such a product possesses properties typical for the
finite dimensional quantum mechanical Hilbert spaces. In particular, 
\begin{equation}
<\varphi _{p}\mid \psi _{q}>=C\delta _{p,q}\text{ and }<\varphi _{p}\mid
\varphi _{p}>>0,  \tag{63}
\end{equation}%
where $C$ is some positive constant. With respect to such defined scalar
product it is possible to define all conjugate operators, e.g. $d^{\ast }$,
etc. and, most importantly, the Laplacians 
\begin{align}
\Delta & =dd^{\ast }+d^{\ast }d,  \notag \\
\square & =\partial \partial ^{\ast }+\partial ^{\ast }\partial ,  \tag{64}
\\
\bar{\square}& =\bar{\partial}\bar{\partial}^{\ast }+\bar{\partial}^{\ast }%
\bar{\partial}.  \notag
\end{align}%
All this was known to mathematicians before Witten's work, Ref.[29\textbf{].}
The unexpected twist occurred when Witten suggested to extend the notion of
the exterior derivative $d$. Within the de Rham picture (valid for both real
and complex manifolds) let $M$ be a compact Riemannian manifold and $K$ be
the Killing vector field which is just one of the generators of isometry of $%
M,$ then Witten suggested to replace the exterior derivative operator $d$ by
the extended operator 
\begin{equation}
d_{s}=d+si(K)  \tag{65}
\end{equation}%
briefly discussed earlier in the context of the equivariant cohomology. Here 
$s$ is real nonzero parameter conveniently chosen. Witten argues that one
can construct the Laplacian (the Hamiltonian in his formulation) $\Delta $
by replacing $\Delta $ by $\Delta _{s}=d_{s}d_{s}^{\ast }+d_{s}^{\ast }d_{s}$
. This is possible if and only if $d_{s}^{2}=d_{s}^{\ast 2}$ $=0$ or, since $%
d_{s}^{2}=s\mathcal{L}(K)$ , where $\mathcal{L}(K)$ is the Lie derivative
along the field $K$, if the Lie derivative acting on the corresponding
differential form vanishes. The details are beautifully explained in the
much earlier paper by Frankel, Ref.[62]. Atiyah and Bott\ observed that the
auxiliary parameter \textbf{s} can be identified with earlier introduced 
\textbf{f}. This observation provides the link between the D-H formalism
discussed earlier and Witten's supersymmetric quantum mechanics.

Looking at Eq.s (64) and following Ref.s[14\textbf{,}51\textbf{,}58] we
consider the (Dirac) operator $\acute{\partial}=\bar{\partial}+\bar{\partial}%
^{\ast }$ and its adjoint with respect to scalar product, Eq.(62). Then use
of above references suggests that the dimension $Q$ of the quatum Hilbert
space associated with the reduced phase space of the D-H integral considered
earlier is given by 
\begin{equation}
Q=\ker \acute{\partial}-co\ker \acute{\partial}^{\ast }.  \tag{66}
\end{equation}%
Such a definition was\ also used by Vergne[14]. In view of the results of
the previous section, and, in accord with Ref.[14], we make an
identification: $Q=\mathfrak{P}(k,n).$

We would like to arrive at this result using different set of arguments. To
this purpose we notice first that according to Theorem 4.7. by Wells,
Ref.[61], we have $\Delta =2\square =2\bar{\square}$ with respect to the K%
\"{a}hler metric on $X$. Next, according to the Corollary 4.11. of the same
reference $\Delta $ commutes with $d,d^{\ast },\partial ,\partial ^{\ast },%
\bar{\partial}$ and $\bar{\partial}^{\ast }.$ From these facts it follows
immediately that if we, in accord with Witten, choose $\Delta $ as our
Hamiltonian, then the supercharges can be selected as $Q^{+}=d+d^{\ast }$
and $Q^{-}=i\left( d-d^{\ast }\right) .$ Evidently, this is not the only
choice as Witten also indicates. If the Hamiltonian \ H is acting in \textit{%
finite} dimensional Hilbert space one may require axiomatically that : a)
there is a vacuum state (or states) $\mid \alpha >$ such that H$\mid \alpha
>=0$ (i.e. this state is the harmonic differential form) and $Q^{+}\mid
\alpha >=Q^{-}\mid \alpha >=0$ . This implies, of course, that [H,$Q^{+}]=[$%
H,$Q^{-}]=0.$ Finally, \ once again, following Witten, we may require that $%
\left( Q^{+}\right) ^{2}=\left( Q^{-}\right) ^{2}=$H. Then, the equivariant
extension, Eq.(65), leads to $\left( Q_{s}^{+}\right) ^{2}=$ H+$2is\mathcal{L%
}$($K$). Fortunately, the above \ supersymmetry algebra can be extended. As
it is mentioned in Ref.[61], there are operators acting on differential
forms living on K\"{a}hler (or Hodge) manifolds whose commutators are
isomorphic to $sl_{2}(\mathbf{C})$ Lie algebra. It is known [63] that 
\textit{all} semisimple Lie algebras are made of copies of $sl_{2}(\mathbf{C}%
)$. Now we can exploit these observations using the Lefschetz isomorphism
theorem whose exact formulation is given as Theorem 3.12 in the book by
Wells, Ref.[61]. We are only using some parts of this theorem in our work.

In particular, using notations of this reference we introduce the operator $L
$ commuting with $\Delta $ and its adjoint $L^{\ast }\equiv \Lambda $ .\ It
can be shown, Ref.[61], p.159, that $L^{\ast }=w\ast L\ast $ where, as
before, $\ast $ denotes the Hodge star operator and the operator $w$ can be
formally defined through the relation $\ast \ast =w$, Ref.[61] p.156. From
these definitions it should be clear that $L^{\ast }$ also commutes with $%
\Delta $ on the space of harmonic differential forms (in accord with p.195
of [61]). As part of the preparation for proving of the Lefschetz
isomorphism theorem, it can be shown [61], that 
\begin{equation}
\lbrack \Lambda ,L]=B\text{ and }[B,\Lambda ]=2\Lambda \text{, }[B,L]=-2L. 
\tag{67}
\end{equation}%
At the same time, the Jacobson-Morozov theorem, Ref.[36], and \ results of
Ref.[63], p.37, essentially guarantee that any $sl_{2}(\mathbf{C})$ Lie
algebra can be brought into form 
\begin{equation}
\lbrack h_{\alpha },e_{\alpha }]=2e_{\alpha }\text{ , }[h_{\alpha
},f_{\alpha }]=-2f_{\alpha }\text{ , \ }[e_{\alpha },f_{\alpha }]=h_{\alpha
}\   \tag{68}
\end{equation}%
upon appropriate rescaling. The index $\alpha $ counts thenumber of $sl_{2}(%
\mathbf{C})$ algebras in a semisimple Lie algebra. Comparison between the
above two expressions leads to the Lie algebra endomorphism, i.e. the
operators $h_{\alpha },f_{\alpha }$ and $e_{\alpha }$ act on the vector
space $\{v\}$ to be described below while the operators $\Lambda ,L$ and $B$
obeying the same commutation relations act on the space of differential
forms. It is possible to bring Eq.s(67) and (68) to even closer
correspondence. To this purpose, following Dixmier [64], Ch-r 8, we
introduce operators $h=\sum\nolimits_{\alpha }a_{\alpha }h_{\alpha }$, $%
e=\sum\nolimits_{\alpha }b_{\alpha }e_{\alpha }$, $f=\sum\nolimits_{\alpha
}c_{\alpha }f_{\alpha }.$ Then, provided that the constants are subject to
constraint: $b_{\alpha }c_{\alpha }=a_{\alpha }$ , the commutation relations
between the operators $h$, $e$ and $f$ are \textit{exactly the same} as for $%
B$, $\Lambda $ and $L$ respectively. To avoid unnecessary complications, we
choose $a_{\alpha }=b_{\alpha }=c_{\alpha }=1$.

Next, following Serre, Ref.[35], Ch-r 4, we need to introduce the notion of
the \textit{primitive} vector (or element).This is the vector $v$ such that $%
hv$=$\lambda v$ but $ev=0.$ The number $\lambda $ is the weight of the
module $V^{\lambda }=\{v\in V\mid hv$=$\lambda v\}.$ If the vector space is 
\textit{finite dimensional}, then $V=\sum\nolimits_{\lambda }V^{\lambda }$.
Moreover, only if $V^{\lambda }$ is finite dimensional it is straightforward
to prove that the primitive element does exist. The proof is based on the
observation that if $x$ is the eigenvector of $h$ with weight $\lambda ,$
then $ex$ is also the eigenvector of $h$ with eigenvalue $\lambda -2,$ etc.
Moreover, from the book by Kac [65], Chr.3, it follows that if $\lambda $ is
the weight of $V,$ then $\lambda -<\lambda ,\alpha _{i}^{\vee }>\alpha _{i}$
is also the weight with the same multiplicity, provided that $<\lambda
,\alpha _{i}^{\vee }>\in \mathbf{Z}$\textbf{. }Kac therefore introduces
another module: $U=\sum\nolimits_{k\in \mathbf{Z}}$ $V^{\lambda +k\alpha
_{i}}$. Such a module is finite for finite Weyl-Coxeter reflection groups
and is infinite for the affine reflection groups associated with the affine
Kac-Moody Lie algebras.

We would like to argue that for our purposes it is sufficient to use only 
\textit{finite} reflection (or pseudo-reflection) groups. It should be
clear, however, from reading the book by Kac that the infinite dimensional
version of the module $U$ leads straightforwardly to all known
string-theoretic results. In the case of CFT this is essential, but for
calculation of the Veneziano-like amplitudes this is \textit{not} essential
as we are about to demonstrate. Indeed, by accepting \ the traditional
option we\ \ loose at once our connections with the Lefschetz isomorphism
theorem ( relying heavily on the existence of primitive elements) and with
the Hodge theory in its standard form on which our arguments are based. The
infinite dimensional extensions of the Hodge-de Rham theory involving loop
groups, etc. relevant for CFT can be found in Ref.[66]. Fortunately, they
are not needed for our calculations. Hence, below we work only with the
finite dimensional spaces.

In particular, let $v$ be a primitive element of weight $\lambda $ then,
following Serre, we let $v_{n}=\frac{1}{n!}e^{n}v$ for $n\geq 0$ and $%
v_{-1}=0,$ so that 
\begin{align}
hv_{n}& =(\lambda -2n)v_{n}  \tag{69} \\
ev_{n}& =(n+1)v_{n+1}  \notag \\
fv_{n}& =(\lambda -n+1)v_{n-1}.  \notag
\end{align}%
Clearly, the operators $e$ and $f$ are the creation and the annihilation
operators according to the existing in physics terminology while the vector $%
v$ can be interpreted as the vacuum state vector. The question arises: how
this vector is related to the earlier introduced vector $\mid \alpha >?$
Before providing an answer to this question we need, following Serre, to
settle the related issue. In particular, we can either: a) assume that for
all $n\geq 0$ the first of Eq.s(69) has solutions and all vectors $%
v,v_{1},v_{2}$ , ...., are linearly independent or b) beginning from some $%
m+1\geq 0,$ all vectors $v_{n\text{ }}$are zero, i.e. $v_{m}\neq 0$ but $%
v_{m+1}=0.$ The first option leads to the infinite dimensional
representations associated with Kac-Moody affine algebras just mentioned.
The second option leads to the finite dimensional representations and to the
requirement $\lambda =m$ with $m$ being an integer. Following Serre, this
observation can be exploited further thus leading us to crucial physical
identifications. Serre observes that with respect to $n=0$ Eq.s(69) possess
a (\textquotedblright super\textquotedblright )symmetry. That is the linear
mappings 
\begin{equation}
e^{m}:V^{m}\rightarrow V^{-m}\text{ and \ }f^{m}:V^{-m}\rightarrow V^{m} 
\tag{70}
\end{equation}%
are isomorphisms and the dimensionality of $V^{m}$ and $V^{-m}$ are the
same. Serre provides an operator (the analog of Witten's $F$ operator) $%
\theta =\exp (f)\exp (e)\exp (-f)$ such that $\theta \cdot f=-e\cdot \theta $%
, $\theta \cdot e=-\theta \cdot f$ and $\theta \cdot h=-h\cdot \theta .$ In
view of such an operator, it is convenient to redefine $h$ operator : $%
h\rightarrow \hat{h}=h-\lambda $. Then, for such redefined operator the
vacuum state is just $v$. Since both $L$ and $L^{\ast }=\Lambda $ commute
with the supersymmetric Hamiltonian H and, because of the group
endomorphism, we conclude that the vacuum state $\mid \alpha >$ for H
corresponds to the primitive state vector $v$.

Now we are ready to apply yet another isomorphism following Ginzburg [36],
Ch-r. 4, pp 205-206 \footnote{%
Unfortunately, the original sourse contains absolutely minor mistakes. These
are easily correctable. The corrected results are given in the text.}. To
this purpose we make the following identification 
\begin{equation}
e_{i}\rightarrow t_{i+1}\frac{\partial }{\partial t_{i}}\text{ , }%
f_{i}\rightarrow t_{i}\frac{\partial }{\partial t_{i+1}}\text{ , }%
h_{i}\rightarrow 2\left( t_{i+1}\frac{\partial }{\partial t_{i+1}}-t_{i}%
\frac{\partial }{\partial t_{i}}\right) ,  \tag{71}
\end{equation}%
$i=0,...,m.$ Such operators are acting on the vector space made of monomials
of the type 
\begin{equation}
v_{n}\rightarrow \mathcal{F}_{n}=\frac{1}{n_{0}!n_{1}!\cdot \cdot \cdot
n_{k}!}t_{0}^{n_{0}}\cdot \cdot \cdot t_{k}^{n_{k}},  \tag{72}
\end{equation}%
where $n_{0}+...+n_{k}=n$ . This result is useful to compare with Eq.(61).

Eq.s (69) have now their analogs 
\begin{align}
h_{i}\ast \mathcal{F}_{n}(i)& =2(n_{i+1}-n_{i})\mathcal{F}_{n}(i),  \notag \\
e_{i}\ast \mathcal{F}_{n}(i)& =2n_{i}\mathcal{F}_{n}(i+1),  \tag{73} \\
\text{ }f_{i}\ast \mathcal{F}_{n}(i)& =2n_{i+1}\mathcal{F}_{n}(i-1),  \notag
\end{align}%
where, clearly, one should make the following consistent identifications: $%
m(i)-2n(i)=2\left( n_{i+1}-n_{i}\right) $ , $2n_{i}=n(i)+1$ and $%
m(i)-n(i)+1=2n_{i+1}.$ Next, we define the total Hamiltonian: $h=$ $%
\sum\nolimits_{i=0}^{k}h_{i}$ \footnote{%
The physical meaning of $h$ is discussed in some detail in our earlier work,
Ref.[11].}so that$\sum\nolimits_{i=0}^{k}m(i)=n,$ and then consider its
action on one of the wave functions of the type given by Eq.(72). Since the
operators defined by Eq.s(71) by design preserve the total degree of
monomials of the type given by Eq.(72) (that is they preserve the Veneziano
energy-momentum codition), we obtain the ground state degeneracy equal to $%
\mathfrak{P}(k,n)$ in agreement with Vergne, Ref.[14], where it was obtained
using different\ methods. Clearly, the factor $\mathfrak{P}(k,n)$ is just
the number of solutions in nonnegative integers to $n_{0}+...+n_{k}=n$,
Ref.[53], p. 252.

\bigskip

\pagebreak

\bigskip

\textbf{References}

\bigskip

[1] \ G.Veneziano, \ Construction of crossing symmetric, Regge

\ \ \ \ \ behaved, amplitude for linearly rising trajectories,

\ \ \ \ \ \textit{Il Nuovo Chimento} \textbf{57}A (1968) 190-197.

[2] \ M.Green, J.Schwarz, E.Witten, \textit{Superstring Theory}, vol.1,

\ \ \ \ \ (Cambridge U.Press, Cambridge, UK, 1987).

[3] \ R.Stanley, \textit{Combinatorics and Commutative Algebra,}

\ \ \ \ \ \ Birkh\"{a}user, Boston, MA, 1996.

[4] \ S.Chowla,\ A.Selberg, \ On Epstein's Zeta function,

\ \ \ \ \ \textit{J. Reine Angew.Math.} \textbf{227} (1967) 86-100.

[5] \ A.Weil, Abelian varieties and Hodge ring, \textit{Collected Works},

\ \ \ \ \ vol.3, Springer-Verlag, Berlin, 1979.

[6] \ A.Weil, Sur les periods des integrales Abeliennes,

\ \ \ \ \ \textit{Comm.Pure Appl.\ Math}. \textbf{29} (1976) 81-819.

[7] \ B.Gross, On periods of Abelian integrals and formula of

\ \ \ \ \ Chowla and Selberg, \textit{Inv.Math}.\textbf{45} (1978) 193-211.

[8] \ S.Lang, \textit{Introduction to Algebraic and Abelian Functions,}

\ \ \ \ \ Springer-Verlag, Berlin, 1982.

[9] \ A.Kholodenko, New string amplitudes from old Fermat

\ \ \ \ \ (hyper)surfaces, \textit{IJMP} A\textbf{19} (2004) 1655-1703.

[10] A.Kholodenko, New strings for old Veneziano amplitudes I.

\ \ \ \ \ \ Analytical treatment, \textit{J.Geom.Phys}. \textbf{55} (2005)
50-74,

\ \ \ \ \ \ arXiv: hep-th/0410242.

[11] A.Kholodenko, New strings for old Veneziano amplitudes II.

\ \ \ \ \ \ Group-theoretic treatment, \textit{J.Geom.Phys}. (2005) in press,

\ \ \ \ \ \ arXiv: hep-th/0411241.

[12] P.Deligne, Hodge cycles and Abelian varieties,

\ \ \ \ \ \ \textit{LNM}. \textbf{900} (1982) 9-100.

[13] J.Carlson, S.Muller-Stach, C.Peters, \textit{Period Mapping}

\ \ \ \ \ \ \ \textit{and Period Domains} Cambridge U.Press, Cambridge, UK,
2003.

[14] \ M.Vergne, Convex polytopes and quantization of symplectic manifolds,

\ \ \ \ \ \textit{\ \ PNAS} \textbf{93} (1996) 14238-14242.

[15] \ S.Donnachie,G.Dosch, P.Landshoff,O.Nachtmann,

\ \ \ \ \ \ \ \textit{Pomeron Physics and QCD},

\ \ \ \ \ \ \ Cambridge U.Press, Cambridge, UK, 2002.

[16] \ P.Collins, \textit{An Introduction to Regge Theory and High Energy
Physics},

\ \ \ \ \ \ \ Cambridge U.Press, Cambridge, UK, 1977.

[17] \ V.Gribov, \textit{The Theory of Complex Angular Momenta},

\ \ \ \ \ \ \ Cambridge U.Press, Cambridge, UK, 2003.

[18] \ J.Forshaw, D.Ross, \textit{Quantum Chromodynamics and the Pomeron},

\ \ \ \ \ \ \ Cambridge U.Press, Cambridge, UK, 1997.

[19] \ A.Kholodenko, E.Ballard, From Ginzburg-Landau to Hilbert Einstein

\ \ \ \ \ \ \ via Yamabe, Reviews in Mathematical Physics (2006) to be
published,

\ \ \ \ \ \ \ arxiv: gr-qc/0410029.

[20] \ M.Herrero, The standard model, arxiv: hep-ph/9812242.

[21] \ C.Ewerz, The odderon in quantum chromodynamics, arxiv: hep-ph/0306137.

[22] \ V.Batyrev, Variation of the mixed Hodge structure of affine

\ \ \ \ \ \ \ hypersurfaces in algebraic tori, Duke Math.J.\textbf{69}
(1993) 349-409.

[23] \ V.Batyrev, Dual polyhedra and mirror symmetry for Calabi-Yau

\ \ \ \ \ \ \ hypersurfaces in toric varieties, J.Alg.Geom. \textbf{3}
(1994) 493-535.

[24] \ T.Hibi, Dual polytopes of rational convex polytopes,

\ \ \ \ \ \ \ Combinatorica \textbf{12} (1992) 237-240.

[25] \ \ V.De Alfaro, S.Fubini,G.Furlan, C.Rossetti, \textit{Currents in
Hadron Physics},

\ \ \ \ \ \ \ Elsevier Publishing Co. Amsterdam, 1973.

[26] \ M.Audin, \textit{Torus Actions on Symplectic Manifolds,}

\ \ \ \ \ \ \ \ Birkh\"{a}user, Boston, MA, 2004.

[27] \ R.Stanley, Invariants of finite groups and their applications to

\ \ \ \ \ \ \ \ combinatorics, \textit{BAMS} \textbf{1} (1979) 475-511.

[28] \ M.Atiyah, R.Bott, The moment map and equivariant cohomology,

\ \ \ \ \ \ \ \ \textit{Topology }\textbf{23} (1984) 1-28.

[29] \ E.Witten, Supersymmetry and Morse theory,

\ \ \ \ \ \ \ \ \textit{J.Diff.Geom}. \textbf{17} (1982) 661-692.

[30] \ W.Fulton, \textit{Introduction to Toric Varieties,}

\ \ \ \ \ \ \ \ Princeton U. Press, Princeton, 1993.

[31] \ E.Witten, Two dimensional gravity and intersection theory on

\ \ \ \ \ \ \ \ moduli space, \textit{Surv.Diff.Geom}.\textbf{1} (1991)
243-310.

[32] \ M.Kontsevich, Intersection theory on the moduli space of curves

\ \ \ \ \ \ \ \ and the matrix Airy function, \textit{Comm.Math.Phys}. 
\textbf{147 }(1992) 1-23.

[33] \ A.Kholodenko, Kontsevich-Witten model from 2+1 gravity: new exact

\ \ \ \ \ \ \ \ combinatorial solution, \textit{J.Geom.Phys}. \textbf{43}
(2002) 45-91.

[34] \ A.Kholodenko, New strings for old Veneziano amplitudes IV.

\ \ \ \ \ \ \ Combinatorial treatment, in preparation.

[35] \ J-P.Serre, \textit{Algebres de Lie Simi-Simples Complexes},

\ \ \ \ \ \ \ Benjamin , Inc. New York, 1966.

[36] \ V.Ginzburg, \textit{Representation Theory and Complex Geometry,}

\ \ \ \ \ \ \ Birkh\"{a}user-Verlag, Boston, 1997.

[37] \ F.Hirzebruch, D.Zagier, \textit{The Atiyah-Singer Theorem and
Elementary}

\ \ \ \ \ \ \ \ \textit{Number Theory}, Publish or Perish, Berkeley, CA ,
1974.

[38] V.Buchstaber, T.Panov, \textit{Torus actions and Their}

\ \ \ \ \ \ \ \textit{Applications in Topology and Combinatorics,}

\ \ \ \ \ \ \ AMS Publishers, Providence, RI, 2002.

[39] \ R.Dias, S.Robins, \ The Ehrhart polynomial of a lattice polytope,

\ \ \ \ \ \textit{\ \ Ann.Math.} \textbf{145} (1997) 503-518.

[40] \ K.Hori, S.Katz, A.Klemm, R.Phadharipande,R.Thomas, C.Vafa,

\ \ \ \ \ \ \ R.Vakil, E.Zaslow, $\mathit{Mirror}$ $\mathit{Symmetry}$, AMS
Publishers,

\ \ \ \ \ \ \ Providence, RI, 2003.

[41] \ D.Cox,S.Katz, \textit{Mirror Symmetry and Algebraic Geometry},

\ \ \ \ \ \ \ AMS Publishers, Providence, RI, 2003.

[42] \ N.Ashcroft, D.Mermin, \textit{Solid State Physics}, Saunders College
Press,

\ \ \ \ \ \ \ Philadelphia, PA, 1976.

[43] \ B.Greene, M.Plesser, Duality in Calabi-Yau moduli space,

\ \ \ \ \ \ \ Nucl.Phys.\textbf{B338} (1990) 15-37.

[44] \ O.Debarre, Fano varieties, in

\ \ \ \ \ \ \textit{Higher Dimensional Varieties and Rational Points, }

\ \ \ \ \ \ \ pp.93-132, Springer-Verlag, Belin, 2003.

[45] \ C.Haase, I.Melnikov, The reflexive dimension of a lattice polytope,

\ \ \ \ \ \ \ arxiv : math.CO/0406485.

[46] \ P.Frampton, \textit{Dual Resonance Models}, W.A.Benjamin, Inc.,

\ \ \ \ \ \ \ Reading, MA, 1974.

[47] \ S.Mandelstam, Veneziano formula with trajectories spaced by two units,

\ \ \ \ \ \ \ Phys.Rev.Lett.\textbf{21} (1968) 1724-1728.

[48] \ \ A.Kholodenko, New strings for old Veneziano amplitudes III.

\ \ \ \ \ \ \ \ Symplectic treatment, J.Geom.Phys.(2005) in press,

\ \ \ \ \ \ \ \ arxiv: hep-th/0502231.

[49] \ A.Barvinok, Computing the volume, counting integral points,

\ \ \ \ \ \ \ \ and exponential sums, \textit{Discr.Comp.Geometry} \textbf{10%
} (1993) 123-141.

[50] \ M.Brion, Points entiers dans les polyedres convexes,

\ \ \ \ \ \ \ \textit{\ Ann.Sci.Ecole Norm. Sup}. \textbf{21} (1988) 653-663.

[51] \ V.Guillemin, V.Ginzburg, Y.Karshon, \textit{Moment Maps,}

\ \ \ \ \ \ \ \textit{Cobordisms, and Hamiltonian Group Actions,}

\ \ \ \ \ \ \ AMS Publishers, Providence, RI, 2002.

[52] \ M.Atiyah, \ Convexity and commuting Hamiltonians,

\ \ \ \ \ \ \ \textit{London Math.Soc.Bull}. \textbf{14} (1982)1-15.

[53] \ M.Vergne, in E.Mezetti, S.Paycha (Eds),

\ \ \ \ \ \ \ \textit{European Women in Mathematics}, pp 225-284,

\ \ \ \ \ \ \ World Scientific, Singapore, 2003.

[54] \ M.Atiyah, Angular \ momentum, convex polyhedra

\ \ \ \ \ \ \ and algebraic geometry,

\ \ \ \ \ \ \ \textit{Proc. Edinburg Math.Soc}. \textbf{26} (1983)121-138.

[55] \ A.Barvinok, \textit{A Course in Convexity},

\ \ \ \ \ \ \ AMS Publishers, Providence, RI, 2002.

[56] \ M.Brion, M.Vergne, Lattice points in simple polytopes,

\ \ \ \ \ \ \ \textit{J.AMS} \textbf{10} (1997) 371-392.

[57] \ A.Khovanskii, A.Pukhlikov, A Riemann--Roch theorem

\ \ \ \ \ \ \ for integrals and sums of quasipolynomials over virtual

\ \ \ \ \ \ \ polytopes, \textit{St.Petersburg Math.J}. \textbf{4} (1992)
789-812.

[58] \ V.Guillemin, \textit{Moment Maps and Combinatorial}

\ \ \ \ \ \ \ \textit{Invariants of Hamiltonian T}$^{n}$\textit{\ Spaces,}

\ \ \ \ \ \ \ Birkh\"{a}user, Boston, MA, 1994.

[59] \ M.Brion, M.Vergne, \ An equivariant Riemann-Roch theorem

\ \ \ \ \ \ \ for complete simplicial toric varieties,

\ \ \ \ \ \ \textit{J.Reine Angew.Math}. \textbf{482} (1997) 67-92.

[60] \ V.Guillemin, S.Sternberg, \textit{Supersymmetry and Equivariant}

\ \ \ \ \ \ \ \textit{de Rham Theory,} Springer-Verlag, Berlin, 1999.

[61] \ R.Wells, \textit{Differential Analysis on Complex Manifolds,}

\ \ \ \ \ \ \ Springer-Verlag, Berlin, 1980.

[62]\ \ T.Frankel, Fixed points and torsion on K\"{a}hler manifolds,

\ \ \ \ \ \ \ Ann.Math.\textbf{70} (1959) 1-8.

[63] \ J.Humphreys, \textit{Introduction to Lie Algebras and Representations}

\ \ \ \ \ \ \ \textit{Theory,} Springer-Verlag, Berlin, 1972.

[64] \ J.Dixmier, \textit{Enveloping Algebras,}

\ \ \ \ \ \ \ Elsevier, Amsterdam, 1977.

[65] \ V.Kac, \textit{Infinite Dimensional Lie Algebras,}

\ \ \ \ \ \ \ Cambridge U. Press, 1990.

[66] \ A.Huckelberry, T.Wurzbacher, \textit{Infinite Dimensional Kahler
Manifolds},

\ \ \ \ \ \ \ \ Birkh\"{a}user, Boston, 1997.

\bigskip

\end{document}